\PassOptionsToPackage{table}{xcolor}
\documentclass[apj, twocolumn, appendixfloats]{openjournal}

\usepackage{macros}

\usepackage[utf8]{inputenc}

\usepackage{orcidlink}
\usepackage{acronym}
\usepackage{graphicx}
\usepackage{amsmath}
\usepackage{float}
\usepackage{appendix}
\usepackage{threeparttable}
\usepackage{gensymb}
\usepackage{array}
\usepackage{multirow}
\usepackage[table]{xcolor}
\usepackage{makecell}

\defcitealias{Dorman2015}{D15}
\defcitealias{Quirk22}{Q22}
\defcitealias{Pessa23}{P23}

\newcommand{\vphio} {$v_{\phi,0}\,$}
\newacro{RGB}{red giant branch}

\hypersetup{
    colorlinks=true,
    linkcolor=blue,
    filecolor=blue,      
    urlcolor=blue,
    citecolor=blue,
}
\usepackage{hyperref}
\begin{document}

\title{\vspace{-0.8 cm}
Stellar Velocity Dispersion versus Age: Consistency across Observations and Simulations, with the Milky Way as an Outlier
\vspace{-1.5 cm}}
%Stellar velocity dispersion and its dependence on age: comparing the Milky Way, M31, M33, PHANGS, and FIRE simulations

%\shorttitle{}
\shortauthors{McCluskey et al.}

\author{Fiona McCluskey$^1$}
\author{Andrew Wetzel\,\orcidlink{0000-0003-0603-8942}$^1$}
\author{Sarah R. Loebman\,\orcidlink{0000-0003-3217-5967}$^2$}
\author{Jorge Moreno\,\orcidlink{0000-0002-3430-3232}$^3$}

%\correspondingauthor{Fiona McCluskey}
\thanks{Corresponding author: Fiona McCluskey}
\email{fmccluskey@ucdavis.edu}

\affiliation{$^1$Department of Physics \& Astronomy, University of California, Davis, CA 95616, USA}
\affiliation{$^2$Department of Physics, University of California Merced, 5200 Lake Road, Merced, CA 95343, USA}
\affiliation{$^3$Department of Physics \& Astronomy, Pomona College, Claremont, CA 91711, USA}

\begin{abstract}
Within disk galaxies, the velocity dispersion, $\sigma$, of stars increases with age, $\tau$, as measured in the Milky Way (MW) and nearby galaxies.
This relation provides a key window into galactic formation history, tracing both the kinematics of stars at birth and the dynamical heating of stars after birth.
We compile and compare observational measurements of the MW, M31, M33, and 16 galaxies from the PHANGS survey.
The MW exhibits significantly colder stellar kinematics, with 2-3 times lower $\sigma(\tau)/v_{\phi,0}$ at a given age, than all but one other observed galaxy.
Therefore, the MW is a kinematic outlier.
To assess how measurement effects influence $\sigma(\tau)$, we analyze the FIRE-2 cosmological simulations, quantifying the impact of uncertainties in stellar age, aperture size, galactocentric radius, and galaxy inclination.
Aperture size and galactocentric radius affect $\sigma(\tau)$ by up to a factor of $\approx2$ for stars younger than 100 Myr, with milder effects on older stars.
Age uncertainties up to 40\% change the \textit{value} of $\sigma(\tau)$ at a given age by $\lesssim20\%$ but can reshape the relation with age and erase merger signatures.
We compare $\sigma(\tau)/v_{\phi,0}$ in FIRE-2 simulations with observations.
FIRE-2 agrees well with M31 and M33 at all measured ages, and with PHANGS for stars older than $\approx500\Myr$.
The average $\sigma(\tau)/v_{\phi,0}$ in FIRE-2 is about two times higher than the MW at most ages, but the youngest stars show better agreement.
The velocity ratios ($\sigma_{\phi}/\sigma_{R}$, $\sigma_{Z}/\sigma_{\phi}$, $\sigma_{Z}/\sigma_{R}$) in FIRE-2 broadly agree with the MW.
We conclude that $\sigma(\tau)$ in FIRE-2, and most cosmological zoom-in simulations, reasonably matches observed nearby galaxies, but matching the MW is rare, because it is a kinematic outlier.
\end{abstract}

\section{Introduction}

Galactic discs form and grow over cosmic time, and the stars within them retain a lasting record of this dynamical evolution. 
Nowhere is this stellar record richer than in the Milky Way (MW), and much work has focused on measuring the velocity dispersion, $\sigma$, as a function of stellar age, $\tau$, in the solar neighborhood.\footnote{While this age velocity dispersion relation is often abbreviated as ``AVR'', we refer to it as $\sigma(\tau)$ throughout this paper.}
Almost all works agree that $\sigma$ typically increases with age, that is, young stars are kinematically cold, exhibiting tidily circular orbits, while older stars are on less coherently circular orbits and therefore have higher $\sigma$. However, despite decades of extensive study, a fundamental question persists: What is the physical origin of $\sigma_{\tau}$? 

The present-day orbits of stars, including $\sigma$ of a population at a given age, reflect two key aspects of a galaxy's formation history: the conditions of the (star-forming) interstellar medium (ISM) over time, which sets the initial orbits of stars at birth, and their subsequent dynamical evolution. 
Both processes affect $\sigma(\tau)$, but their relative contributions remain uncertain.

In the canonical picture of \cite{Spitzer51}, stars form with the same low $\sigma$ at all ages, but repeated scattering interactions with (then undetected) giant molecular clouds (GMCs) gradually increase $\sigma$ as a stellar population ages.
However, analytical models of GMC-driven heating failed to match observation, motivating the study and acceptance of additional dynamical heating processes, such as transient spiral arms \citep{Barbanis67, Carlberg85, Minchev06}, bars \citep{Saha10}, perturbations by satellite galaxies \citep{Bird12, D'Onghia16, Carr22}, and galaxy mergers \citep{Abadi03, Villalobos08}.
The current consensus is that perturbations from transient spiral structure efficiently increase the in-plane $\sigma$, while GMC scattering acts more isotropically and redirects motions out of the plane \citep{bt-08, Sellwood13}.
Yet, standard heating models fail to reach the high $\sigma$ observed for the MW's older stars.
This discrepancy is unsurprising, because traditional models assume that the MW's present-day conditions are representative of its earlier states, such that stars formed with the same low $\sigma$ over cosmic time and subsequently experienced the same dynamical heating processes.

In contrast, nearly all recent observations and theoretical models of galaxy formation find that the dynamical state of the ISM evolves substantially over cosmic time.
At higher redshifts ($z \gtrsim 1$), disk galaxies are consistently thicker, clumpier, and more turbulent than local disks \citep[for example][]{schreiber-09, Birkin23}, probably due to their higher gas fractions and star-formation, accretion, and merger rates \citep{Conselice05, Swinbank12, Tacconi13, Genzel17, Stott16, FS20}.
%At $z \gtrsim 1$, disk galaxies were consistently thicker, clumpier, and more turbulent than disks today \citep[for example][]{schreiber-09, Birkin23}, likely from higher gas fractions, higher rates of gas accretion and mergers, and higher rates of star formation and thus feedback \citep[for example][]{Conselice05, Swinbank12, Tacconi13, Genzel17, Stott16, FS20}.
Spatially resolved studies of ionized gas kinematics find that $v_{\phi} / \sigma$, which measures the degree of rotational-support or diskiness, is often as high as $10 - 20$ in the local Universe but decreases with increased redshift, with average values of $\approx 2$ at $z = 3$ \citep[see recent compilation in][]{Danhaive25}.
Similarly, $\sigma$ increases from $\approx 20 \kms$ in disk galaxies today to $\approx 50 \kms$ at $z = 3$ \citep{Kassin-12, Wisnioski15, Wisnioski19, Simons16, Ubler24}.
These works indicate that disks ``settled'' over cosmic time, such that their gaseous disks became progressively less turbulent and more rotationally supported.
Theoretical models, including cosmological simulations, strongly support these observations, generally finding that the progenitors of MW-mass galaxies typically formed long-lived disks at $z \lesssim 1 - 2$, with such disks growing radially ``inside-out'' and vertically ``upside-down'' \citep[for example][]{Bird12, Pillepich19}.

This observational picture indicates that stars form with different dynamics over cosmic time, given that stars inherit their initial kinematics from their progenitor gas. That said, the same conditions that drive increased ISM turbulence -- including substantial gas fractions, higher merger rates, and burstier star formation --  also affect post-formation dynamical heating. For example, feedback-driven outflows and the resulting potential fluctuations can inject additional energy into the stellar component, leading to rapid, non-adiabatic heating that further amplifies stellar velocity dispersions \citep{el-badry18}.
Such ISM conditions also increase the prevalence of scattering from long-lived massive gas clumps and GMCs \citep[for example][]{Bournard14,Johnson18,Price20}. 
Disentangling the degree to which present-day velocity dispersions reflect hotter birth conditions versus cumulative dynamical heating is challenging, as both processes can yield similar observational signatures. For example, \citet{Ting19} argue that the MW's vertical $\sigma(\tau)$ is consistent with stars forming with a constant, low $\sigma$ over the past 8 Gyr, but experiencing stronger GMC-driven dynamical heating at these earlier times. 

This already tangled picture has recently been further complicated, as some recent observations now challenge our understanding of (gaseous) disk evolution. ALMA has revealed cold, rotationally supported molecular gas discs out to $z \sim 7$, including systems with $v/\sigma_0 \gtrsim 4$ \citep{Rizzo20, Lelli21, Pope23}, and even as high as $v/\sigma_0 \sim 11$ at $z = 7.3$ \citep{Rowland24}. JWST now extends this picture to ionized gas, detecting kinematically cold discs across a broad range of stellar masses and star formation rates \citep{Nelson23, Nelson24, Tohill24, Robertson23, deGraaff24a}. Although these findings suggest that dynamically cold disks can form, at least temporarily, in gas, an open question is how representative these disks are at early times. For example, \cite{Danhaive25} showed that almost all of these early disks are massive, $\Mstar > 10^{10} \Msun$, and thus not representative of a typical MW-mass progenitor at these redshifts.

A key limitation of high-redshift studies is that they generally only provide a statistical view of galaxy populations and cannot reveal how any individual galaxy continuously evolved across cosmic time. 
%Yet, the substantial scatter in galaxy properties at a given redshift indicates that MW-progenitors experienced a wide variety of star-formation and accretion histories.
In contrast, detailed measurements of resolved stars (or stellar populations) in nearby galaxies can provide a view of the entire formation history of a galaxy.
In particular, the advent of Gaia \citep{Gaia16}, along with large spectroscopic surveys, such as APOGEE \citep{APOGEE17}, GALAH \citep{GALAH15, GALAH18}, and LAMOST \citep{LAMOST12}, now provides up to 6D orbital phase-space coordinates and elemental abundances for millions of stars throughout the Galaxy. 
In parallel, the growth of asteroseismology \citep{Pinsonneault18, Miglio21}, in combination with high-quality spectroscopic measurements and data-driven models \citep{Ho17, Ness16, Sharma18, Queiroz23}, has driven remarkable progress in our ability to measure the ages of stars.
Such progress enables measurements of $\sigma(\tau)$, which was historically limited to the immediate solar neighborhood and beholden to unclear systematics and age uncertainties \citep{Nordstrom2004, Holmberg07, Casagrande11}, to large swathes across the disk with unprecedented precision \citep[for example][]{Lagarde21, Anders23, Kordopatis23}.

These advancements have revolutionized our view of the MW's formation history. Indeed, a general picture has emerged that the MW began as a turbulent galaxy, subject to a bombardment of mergers and bursts of star formation \citep{Myeong19, Rix22, Chandra23}. It then rapidly ``spun up'', forming a coherently rotating disk $\approx 12 \Gyr$ ago \citep{Belokurov22, Conroy22, Xiang22, Xiang25}. The young disk then faced, and ultimately survived, a head-on collision with a massive satellite $\approx 8 - 11 \Gyr$ ago: this event, the ``Gaia-Sausage-Enceladus'' (GSE) merger \citep{Belokurov18, Helmi18}, scrambled and ``splashed'' the pre-existing disk, but also brought in fresh fuel for star formation and potentially seeded the formation of the thin disk \citep[for example]{Belokurov20, Grand20}. While many aspects of the GSE merger, including its impact on the MW's disk and $\sigma(\tau)$, remain unclear \citep[for example][]{DiMatteo19, Fernandez25, Skuladottir25}, it is generally agreed to be the MW's last major merger to date \citep[though see][about the timing of this event]{Donlon24}. 

Furthermore, there is growing consensus that certain aspects of the MW's formation history -- namely its early disk formation and quiescent merger history since z$\sim$2 -- are atypical among present-day disk galaxies \citep{Hammer07, Evans20, Mackereth18, Kruijssen19b, Semenov24_mw1}. Given how much of our understanding of galaxy formation, and in particular post-formation dynamical heating, is benchmarked against the MW, it is imperative to understand whether the MW is indeed an outlier or simply appears so due to observational bias remains an open question. Because we view it from the inside, we cannot measure it using the same methods applied to other galaxies, complicating efforts to draw meaningful comparisons \citep[though see][]{Fielder21,Wang24}.

%if apparent differences stem from the dramatically different ways we measure the MW versus higher redshift galaxies \citep[for example,][]{Davies11, Rizzo24, Lee25}.

One of the best ways to distinguish these possibilities is via spatially resolved measurements of $\sigma(\tau)$ in nearby disk galaxies.
M31 and M33, the only other disk galaxies in the Local Group (LG), provide our best opportunities, motivating surveys by the Hubble Space Telescope (HST) \citep{Dalcanton12, Williams21} and ground-based observatories \citep{Guhathakurta06, McConnachie09, McConnachie18, Bhattacharya19}.
In many ways, M31 is similar to the MW: a slightly more massive disk galaxy with a central bar, spiral arms, and modest star formation \citep{Yin09, Mutch11}.
In other ways, M31 is a foil to the quiescent MW: it likely experienced a major ($\approx 4\mathpunct{:}1$) merger $\approx 2.5 \Gyr$ ago \citep{Hammer18, D'Souza18}, resulting in a thicker overall disk \citep{Dalcanton23, Gibson24}. While M31 allows us to compare the MW to a more massive, recently-interacting galaxy, its most massive satellite, M33, provides a view of a disk in the lower-mass regime. With a stellar mass of $4.8 \times 10^{9} \Msun$, M33 occupies the mass scale at which thin disks emerge \citep[for example][]{Dekel09, Simons16}, and its flocculent spiral structure and recent bursts of star formation provide valuable insights into the dynamical evolution of the ISM and post-formation heating processes \citep{Smercina23,Peltonen24}.
Importantly, observations of M31 \citep{Dorman2015} and M33 \citep{Quirk22} have provided our first view of $\sigma(\tau)$ for a star-forming disk galaxy other than the MW. 

Beyond the LG, integral-field spectroscopy (IFS) provides an intermediate view between galaxy-wide integrated properties and individual stars. While many IFS surveys, including MANGA \citep{Bundy15}, SAMI \citep{Croom12}, and CALIFA \citep{Sanchez12}, strike a middle ground between survey scope and depth, measuring spatially resolved kinematics for up to 10,000 galaxies at intermediate resolution ($0.5 - 1 \kpc$), some surveys focus on higher-resolution ($\lesssim 100 \pc$) measurements for smaller samples.
The PHANGS-MUSE survey \citep{PHANGS-MUSE} obtained stellar kinematics for 19 star-forming galaxies at cloud-scale ($\approx 50 - 100 \pc$) resolution, measuring $\sigma(\tau)$ in disk galaxies outside the LG for the first time \citep{Pessa23}. These results, along with those of other MUSE surveys, including MAD \citep{denBrok20}, TIMER \citep{Gadotti19}, and GECKOS \citep{vandeSande24}, allow us to place the detailed dynamical record of the MW (and the LG) in a greater cosmological context.

Cosmological baryonic simulations are valuable tools for interpreting observations of stellar kinematics, as they now produce realistic disk galaxies spanning a wide range of formation histories \citep[for reviews see][]{Naab17, Vogelsberger20, Crain23}. 
Zoom-in simulations, in particular, contain both cosmological nonequilibrium perturbations like mergers and mass growth, with the resolution needed to model the multiphase ISM, enabling star formation and feedback to be treated self-consistently in a broad context. 
Notably, these simulations resolve both global galaxy properties (e.g., morphology, disk dynamics) and the smaller-scale structures (e.g., GMCs, spiral arms, bars) that drive post-formation dynamical heating \citep{Benincasa20, Yu22, Ansar25}. 
This enables detailed analysis of the evolving strength and balance between birth kinematics and subsequent heating across cosmic time.
By tracing galaxy evolution across time and comparing simulated galaxies with each other and with observations, these models reveal how present-day kinematics reflect both birth conditions and cumulative dynamical processes imprinted on $\sigma(\tau)$.

In this work, we focus on $\sigma(\tau)$ in the MW and nearby disk galaxies of similar mass, using FIRE-2 simulations (described in \S{\ref{sec:method}}) as a comparison baseline. 
Although some of these works \citep[for example][]{Dorman2015} note that the observed $\sigma(\tau)$ in M31 differs from that of the MW, no work has compiled and compared all available measurements of $\sigma(\tau)$ in nearby MW-mass disk galaxies. 
In this work, we aim to (1) put these observations on more equal footing, and (2) provide a cogent  view of the state of the literature.

In \S\ref{sec:P1}, we assess how measurement considerations, including aperture size, galaxy inclination, age uncertainties, and galactocentric radius, affect the inferred $\sigma(\tau)$. 
In \S\ref{sec:P2}, we compile existing observations, focusing exclusively on star-forming disk galaxies; we exclude non-disky low-mass systems such as the LMC, SMC, and other MW satellites \citep[for example][]{Cole05,Olsen07,Leaman17}, as well as lenticular/S0 galaxies \citep{Poci19, Poci21}. We summarize and interpret these results in the discussion (\S\ref{sec:discussion}).

\section{Methods}
\label{sec:method}

\subsection{FIRE-2 Simulations}
\label{subsec:sims}

We study the stellar disks of 11 simulated MW-mass galaxies from the Feedback in Realistic Environments (FIRE) project \footnote{
FIRE project web site: \href{http://fire.northwestern.edu}{http://fire.northwestern.edu}}, including 5 isolated galaxies from the \textit{Latte} suite \citep{Wetzel16} and 3 LG-like pairs from the \textit{ELVIS on FIRE} suite \citep{Garrison-Kimmel18}.
These baryonic cosmological simulations model the physics of stars and gas (in addition to dark matter) in a ``zoom-in'' region embedded within a lower-resolution cosmological background. Crucially, this zoom-in method enables these simulations to reach the resolution needed to resolve the multiphase ISM and GMCs.

These simulations use the zoom-in technique to simulate MW-mass galaxies at high resolution within a fully cosmological context. We first run dark-matter-only cosmological simulations within a periodic volume of side length $70.4 - 172 \Mpc$, assuming a flat $\Lambda$CDM cosmology with parameters broadly consistent with \citet{planck20}: $h = 0.68 - 0.71$, $\Omega_{\Lambda} = 0.69 - 0.734$, $\Omega_{\rm m} = 0.266 - 0.31$, $\Omega_{\rm b} = 0.0455 - 0.048$, $\sigma_{8} = 0.801 - 0.82$ and $n_{\rm s} = 0.961 - 0.97$.
After selecting a region centered on a halo (or pair of halos) of interest at $z = 0$, we use \textsc{MUSIC} \citep{hahn11} to identify the particles within the region, trace them back to $z \approx 99$, and regenerate the initial conditions for that region at higher resolution and including baryons.

For the \textit{Latte} suite, halo selection is based solely on their mass at $z = 0$ ($M_{\rm 200m} \approx 1 - 2 \times 10^{12} \Msun$) and relative isolation from similarly massive halos.
The \textit{ELVIS on FIRE} suite selection is a pair of neighboring ($d < 600 - 1000 \kpc$), massive ($M_{\rm 200m} \approx 1 - 3 \times 10^{12} \Msun$ halos approaching each other, to mimic the LG.
We emphasize that the selection of these halos was agnostic to any halo properties beyond mass (and LG-like environment), including formation history, concentration, spin, or satellite population.
That is, there was no attempt to recreate the MW, or any other individual galaxy, or the larger-scale environment (beyond the LG).
We exclude three galaxies (m12r, m12w, and m12z) from our analysis, because they lack coherent (rotationally-dominated) stellar disks, making them poor analogs for the observed disk galaxies we compare against.

We ran the simulations using GIZMO\footnote{\url{http://www.tapir.caltech.edu/~phopkins/Site/GIZMO.html}}, a multimethod gravity plus (magneto)hydrodynamics code, in its meshless finite-mass (MFM) mode.
This mesh-free Lagrangian Godunov method combines the advantages of smoothed-particle hydrodynamics methods and grid-based adaptive mesh refinement schemes while avoiding many of their drawbacks: this method allows automatically adaptive resolution, accurate gravitational coupling, and reduced advection errors, all while accurately capturing instabilities and shocks and maintaining exact mass, energy, and momentum conservation and excellent angular momentum conservation \citep[for extensive tests see][]{Hopkins15, HopkinsRaives16, Hopkins17}. For gravity, GIZMO couples hydrodynamics to an improved version of the $N$-body TreePM solver from GADGET-3 \citep{Springel05} using fully adaptive and conservative gravitational force softenings, so that the gravitational force softening for gas always matches the hydrodynamic smoothing length, following \cite{Price07}.

The \textit{Latte} suite has a dark-matter particle mass of $3.5 \times 10^5 \Msun$ and an initial baryon particle mass of $7070 \Msun$, although stellar mass loss causes star particles to have typical masses $\approx 5000 \Msun$ today.
Gas cells have fully adaptive gravitational softenings that match the hydrodynamic kernel smoothing: softening lengths are $\approx 20 - 40 \pc$ at typical ISM densities ($n \sim 1 \rm cm^{-3}$) but reach $\approx 1 \pc$ in the densest regions. In contrast, dark-matter and star particles have fixed gravitational force softenings, comoving at $z > 9$ and physical thereafter, with a Plummer equivalent of $\epsilon_{\rm dm} = 40 \pc$ and $\epsilon_{\rm star} = 4 \pc$ (corresponding to the typical interparticle spacing in star-forming gas).
The \textit{ELVIS on FIRE} suite has a mass resolution $\sim 2 \times$ better than the \textit{Latte} suite: Romeo \& Juliet have initial baryon particle masses of $3500 \Msun$, while Romulus \& Remus and Thelma \& Louise have $4000 \Msun$.

The simulations incorporate metallicity-dependent radiative cooling and heating processes for gas across temperatures $10 - 10^{10} \K$, including free-free, photoionization and recombination, Compton, photoelectric and dust collisional, cosmic ray, molecular, metal-line, and fine structure processes.
Combined with their high resolution, the FIRE-2 simulations can resolve key phase structures in the ISM, allowing gas to collapse into GMCs \citep{Benincasa20, Guszejnov20}.
The simulations self-consistently generate 11 elements (H, He, C, N, O, Ne, Mg, Si, S, Ca, Fe), including a model for sub-grid mixing/diffusion via turbulence \citep{Hopkins16a, Su17, Escala18}.

Star formation occurs in gas that is dense ($n > 1000$ $\rm cm^{-3}$), self-gravitating \citep[following][]{Hopkins13}, self-shielding and molecular \citep[following][]{KG11}, and Jeans unstable.
If all of these criteria are met, a gas cell probabilistically converts to a star particle in a local gravitational free-fall time. Each star particle represents a single stellar population sampled from a \citet{Kroupa01} initial mass function, with mass and metallicity inherited from its progenitor gas cell. Once formed, star particles follow stellar evolution models that tabulate feedback event rates, luminosities, energies, and mass-loss rates from STARBURST99 v7.0 \citep{Leitherer99}.
FIRE-2 implements all of the major channels for stellar feedback, including core-collapse and white-dwarf (Type Ia) supernovae, stellar winds, radiation pressure, photoionization, and photoelectric heating. FIRE-2 also includes photoionization and heating from a spatially uniform, redshift-dependent UV background \citep{Faucher09} that reionizes the Universe at $z \approx 10$.

\subsection{Selecting Stars and Measuring Velocity Dispersion}
\label{subsec:selection}

We measure all positions and velocities in galactocentric cylindrical coordinates ($R$, $\phi$, $Z$).
To determine the center of each galaxy -- and thus our coordinate system -- we iteratively compute the mean center-of-mass position of all stars using the ``shrinking spheres'' method until the spherical radius drops below $10 \pc$.
We then align our coordinate system with the principal axes of the stellar disk, as defined by the moment-of-inertia tensor of the 25\% youngest stars inside a radius that encloses 90\% of the total stellar mass within 10 kpc.
Although we define our coordinate system using the disk-scale properties of young stars, we measure stellar kinematics on local apertures, which in principle can have distinct alignments relative to other apertures and the galaxy as a whole.
However, we found that our results are insensitive to the specifics of our coordinate system, for example, if we instead assign our axes using just the young stars within that aperture, or if we use gas or different definitions for young stars.

\cite{McCluskey} showed that, in these FIRE-2 simulations, $\sigma$ of cold gas and young stars at formation strongly depend on the spatial scale over which one measures them. Specifically, when measured in apertures of radius $250 \pc$, $\sigma$ of cold gas and young stars at formation are 3 times and 2 times \textit{lower} than when measured across an entire galactocentric annulus.
We now extend this analysis to the present-day $\sigma$ of stars of various ages.

Within each aperture, we compute the velocity dispersion, $\sigma$, for a given velocity component ($v_R$, $v_\phi$, or $v_Z$, as half of the mass-weighted $16^{th}$–$84^{th}$ inter-percentile range, which is equivalent to the standard deviation for normal distribution.
From these we also compute the ``total'' 3-D velocity dispersion, $\sigma_{\rm 3D} \equiv \sqrt{\sigma_R^2 + \sigma_\phi^2 + \sigma_Z^2}$.
We then compute the median $\sigma$ across the aperture at a given galactocentric radius, $R$, where we only include apertures that contain at least seven star particles within our desired age range.
Unless otherwise specified, these are cylindrical apertures of radius $250 \pc$ and vertical height $|Z| < 3 \kpc$ that span an annulus centered at $R = 8 \kpc$.

By default we show the ``total'' $\sigma_{\rm 3D}$, because the individual velocity components generally show trends similar to $\sigma_{\rm 3D}$ (or $\sigma_{\rm LOS}$) and each other.
In some cases we show individual components, either because this allows us to compare the FIRE-2 simulations to the MW, as in Section~\ref{subsec:fire comparison}, or because nontrivial differences exist between different components, as in Section~\ref{subsec:age uncertainties}.

We tested how each of the parameters for apertures impacts our results.
Using the mean $\sigma$ of the apertures across an annulus instead of the median typically produces identical results. 
Similarly, the median (or mean) $\sigma$ across an annulus is largely insensitive to the minimum number of star particles, provided $N_{\rm min} \gtrsim 3$.
That said, $\sigma$ within individual apertures occasionally depends on this threshold, with higher $N_{\rm min}$ restricting our results toward denser apertures, which are generally (but not exclusively) biased towards larger $\sigma$.

For our fiducial parameters ($R = 8 \kpc$, $|Z| < 3 \kpc$, and $r_{\rm ap} = 250 \pc$) the median number of star particles within our age bins centered at $\approx 400 \Myr$, 2 Gyr, and 6 Gyr typically exceeds 100 for the ELVIS simulations and 50 for the Latte simulations (given their differing mass resolutions). The narrow range of our youngest age bin ($0 - 100 \Myr$) can reduce this median to $10 - 20$ particles.
Ultimately, we use a threshold of a minimum of 7 star particles to include an aperture, as a balance against Poisson noise within individual apertures and against small-number statistics and density-biasing across a galactic annulus.

Our results depend on the spatial parameters of the aperture, specifically its size and its galactocentric radius, as we explore in Sections~\ref{subsec:ap_size} and \ref{subsec:rad_selection}.
We also examined the impact of the vertical extent of an aperture, finding that it has negligible effects, given that the majority of stars are close to the disk midplane, in agreement with the tests shown in \cite{McCluskey}.

\subsection{Age Binning}
\label{subsec:methods_age}

In Section~\ref{sec:P1}, we show results for stars in five representative age ranges: $0 - 0.1$, $0.1 - 0.6$, $1.75 - 2.25$, $5.75 - 6.25$, and $10.75 - 11.75 \Gyr$.
We choose these age ranges to sample the full dynamical history of the disk and also to complement observations of resolved stars and stellar populations in nearby galaxies.

The youngest age bin, $0 - 100 \Myr$ allows us to study the kinematics of stars near birth and thus probe the dynamical state of the star-forming ISM.
Although stars inherit the kinematics of their progenitor gas cells, dynamical heating of stars quickly can increase and decouple the (effectively dissipationless) stars from the (dissipational) gas.
%However, such narrowness comes at the cost of lower star counts, rendering our results more susceptible to Poisson noise and the pitfalls of low number statistics. Therefore, we limit our youngest bin to ages $\leq$ 0.1 as a trade-off between sufficiently little dynamical evolution and sufficiently large sample size.
This narrow age range also allows us to compare with observations of young stars in nearby galaxies.

The youngest age bins probed by observations of M31 and M33 have mean ages of 30 and 80 Myr, while \citetalias{Pessa23} presented $\sigma(\tau)$ of stellar populations in PHANGS galaxies with luminosity-weighted ages $< 100 \Myr$.

Our second age bin is $100 - 600 \Myr$.
While still young, these stars have evolved over 1-3 disk dynamical times, so they provide a window into the internal heating processes that shape stellar dynamics in the disk.
This age range also overlaps with the youngest age bins used in MW observations, and it bookends the ages of young AGB stars in M31 (which have a mean age of 400 Myr), and exactly matches the ``intermediate'' age bin of \citetalias{Pessa23} for PHANGS galaxies.

For the three older age bins, we chose age ranges that correspond to the typical midpoints of the disk dynamical eras that we presented for the FIRE-2 simulations in \cite{McCluskey}.
Briefly, \cite{McCluskey} found that the formation of stellar disks -- as traced by the $v_\phi / \sigma$ of newly formed stars -- proceeded in three distinct phases.
Stars in the ``Pre-Disk'' Era formed on non-disk, dispersion-dominated orbits. Next, the ``Early-Disk'' Era commenced once stars began to form (permanently) on rotation-dominated orbits. These early disks were thick and kinematically hot, with high $\sigma$.
Subsequent generations of stars formed on increasingly rotationally dominated orbits, and eventually the galaxies settled into their ``Late-Disk'' Era, forming stars on consistently dynamically cold orbits. Although the exact timing and length of these transitions vary across the sample, these simulations typically transitioned to the Early-Disk Era 8 Gyr ago and the Late-Disk Era 4 Gyr ago, such that stars with ages of $1.75-2.25$, $5.75-6.25$, and $10.75-11.25 \Gyr$ formed predominantly in the Late-, Early-, and Pre-Disk Eras, respectively.

\begin{table}
\caption{
stellar properties today for the FIRE-2 simulations of MW/M31-mass galaxies that we analyze
}
\setlength\extrarowheight{2.3pt}
\centering
\begin{tabular}{|c|c|c|c|c|c|}
\hline
Simulation & $M_{\star, 90}$ & $v_{\rm \phi,0}$ & $\sigma_{\rm Z}$ & $\sigma_{\rm 3D}$ & $\tau_{\rm disk}$ \\
Name & $\mathrm{[10^{10} M_\odot]}$ & [km/s] & [km/s] & [km/s] &  [Gyr] \\
\hline
\hline
m12m & 10.0 & 288 & 13.3 & 33.1 & 9.2 \\
Romulus & 8.0 & 265 &  16.6 & 41.8 & 7.4 \\
m12b & 7.3 & 272 & 13.1 & 37.5 & 7.4 \\
m12f & 6.9 & 256 & 13.5 & 31.6 & 7.4 \\
Thelma & 6.3 & 230 & 19.4 & 43.9 & 4.4  \\
Romeo & 5.9 & 249 & 12.2 & 31.3 & 11.0 \\
m12i & 5.3 & 238 & 12.8 & 27.5 & 6.7  \\
m12c  & 5.1 & 240 & 19.5 & 38.4 & 6.5 \\
%m12w & 4.8 & 163 & 5.5 & 29.8 & 4.1 \\
Remus & 4.0 & 216 & 10.7 & 25.0 & 7.9 \\
Juliet & 3.3 & 211 & 15.0 & 32.6 & 4.4 \\
Louise & 2.3 & 177 & 11.0 & 26.1 & 7.2 \\
%m12z & 1.8 & 92.6 & 19.0 & 39.0 & 0.5 \\
%m12r & 1.5 & 127 & 7.3 & 15.4 & 5.9 \\
\hline
Average & 6.0 & 236 & 13.5 & 32.4 & 7.2 \\
\hline
\end{tabular}
\tablecomments{
We list galaxies by decreasing stellar mass.
The first column lists the galaxy name: `m12' indicates an isolated galaxy; otherwise the galaxy is in a Local Group-like pair.
$M_{\rm star, 90}$ is the stellar mass within $R_{\rm star, 90}$, the radius enclosing 90\% of the stellar mass within 20 kpc.
$v_{\phi,0}$, $\sigma_{Z}$ and $\sigma_{\rm 3D}$ are the median rotational velocity, vertical velocity dispersion, and total 3D velocity dispersion of stars younger than 100 Myr, as measured in cylindrical apertures of radius of $250 \pc$ around an annulus centered at $R = 8 \kpc$.
$\tau_{\rm disk}$ is the lookback time when the galaxy formed a long-lived disk and all stars formed with $v_{\phi} / \sigma_{\rm 3D} > 1$, from \cite{McCluskey}.
Bottom row shows the mean across the 11 galaxies.
}
\label{table:table_fire}
\end{table}

\section{Measurement Effects}
\label{sec:P1}

In this section, we use the FIRE-2 simulations to investigate how uncertainties, systematics, and measurement choices impact the measured stellar velocity dispersions.
We seek to provide guidance for interpreting observations, comparing them against each other, and comparing them against simulations.

\subsection{Age Uncertainties}
\label{subsec:age uncertainties}

Stellar ages are difficult to measure precisely, and their uncertainties are generally $\gtrsim 20 - 30\%$ \citep[for example][]{Soderblom10}.
Fortunately, the synergy of astrometric and spectroscopic surveys with asteroseismological data has driven tremendous progress in the past decade \citep[for example][]{Silva18, Mackereth19}.
Nevertheless, stellar-age uncertainties continue to hinder our ability to accurately reconstruct the formation history of the MW and nearby galaxies, both by obscuring merger features \citep{Martig14, Buck20} and by decreasing inferred dynamical heating \citep{Aumer16}.

We explore the impact of age uncertainties on the measured $\sigma(\tau)$.
To model this, we generate new ages for each star particle, randomly drawing ``measured’’ ages from Gaussian distributions centered on the true age with standard deviations corresponding to 10, 20, 30, and 40\% (fractional) uncertainties.
We then recompute $\sigma$ in each age bin using these ages.

The degree to which age uncertainties influence $\sigma(\tau)$ depends mainly on the steepness and shape of the true dependence on age.
In general, if $\sigma(\tau)$ is smooth and slowly varying, modest age uncertainties do little to distort the overall trend. However, by mixing populations across age bins, uncertainties generally lead to an inferred dependence on age that is shallower than truth.
Furthermore, if $\sigma(\tau)$ exhibits sharp features -- such as spikes or breaks in slope -- age uncertainties can smear these signatures by mixing stars across bins with different intrinsic kinematics. The impact also varies with age: for a fixed fractional age uncertainty, absolute uncertainties are smaller at younger ages.
Accordingly, we expect minimal effects from age errors at young ages, even when the slope is steep, and more noticeable smoothing at older ages, where the population size is limited, absolute uncertainties are larger, and slopes are (occasionally) steep.

\subsubsection{Case Studies}
\label{subsubsec:case studies}

\begin{figure*}
\centering
\includegraphics[width = 0.65 \textwidth]{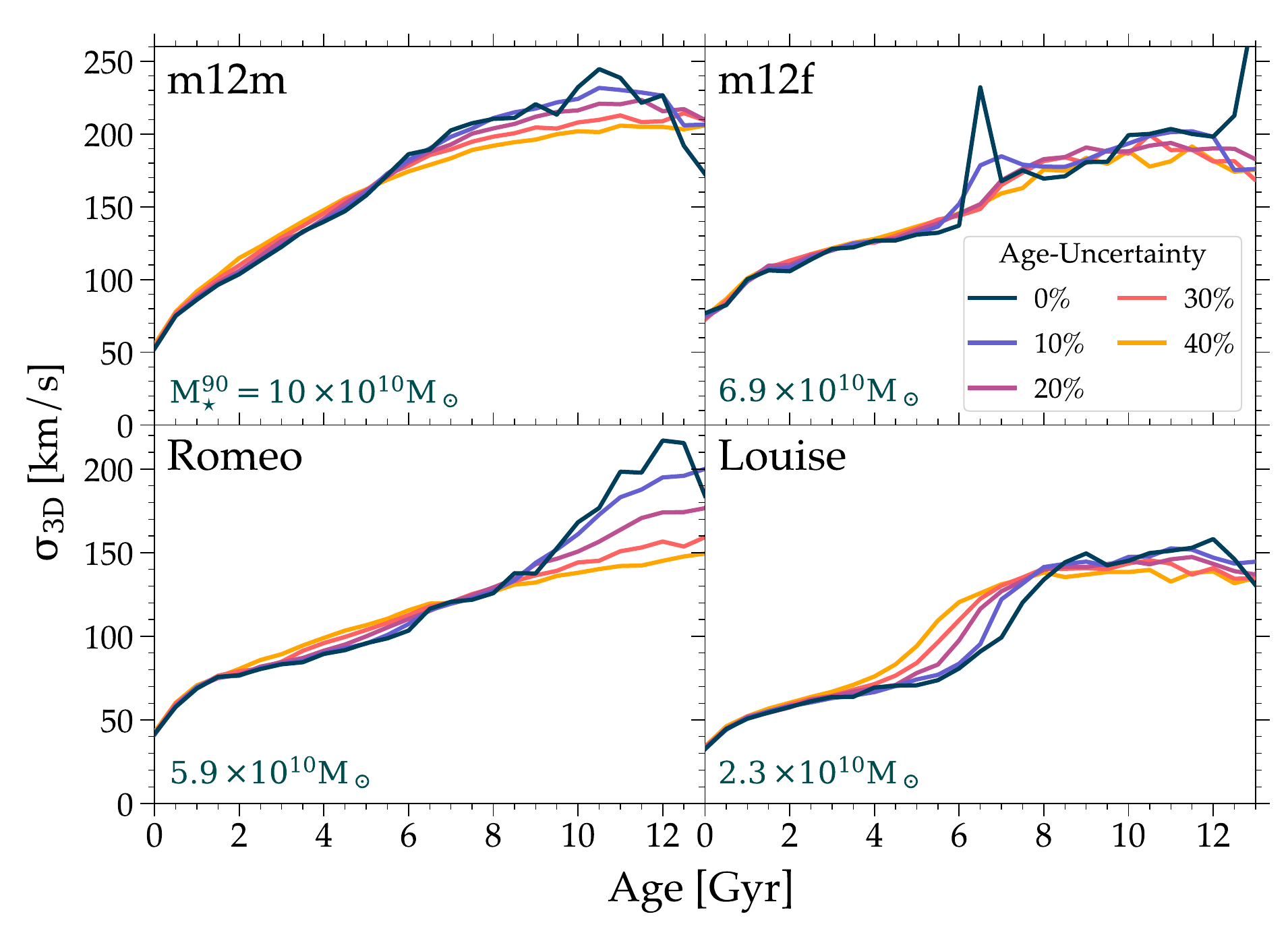}
\vspace{-4 mm}
\caption{
\textbf{Stellar 3D velocity dispersion versus age in 4 representative galaxies}, with $\Mstar^{\rm 90}$ today in each corner.
We measure $\sigma_{\rm 3D}$ at $R = 8 \kpc$, imposing fractional uncertainties on stellar ages of 0\%, 10\%, 20\%, 30\% and 40\%.
$\sigma_{\rm 3D}$ generally increases with age, reflecting the disk settling over time \textit{and} the dynamical heating of stars after birth.
\textbf{m12m (top left)} show a nearly linear relation, largely unaffected by age uncertainties at ages $\lesssim 6 \Gyr$, but the relation flattens at ages $> 6 \Gyr$, with larger uncertainties leading to progressively flatter slopes.
\textbf{m12f (top right)} shows dramatic behavior: a major merger $\approx 7 \Gyr$ ago caused a spike, but age uncertainties $\gtrsim 20\%$ erase this.
\textbf{Romeo (bottom left)} is our closest analog to the MW.
Age uncertainties have mild effects up to $\approx 9 \Gyr$, before which the strong increase with age becomes increasingly flattened for larger uncertainties.
\textbf{Louise (bottom right)} is our lowest-mass galaxy.
The rapid rise in $\sigma_{\rm 3D}$ $\approx 7 \Gyr$ ago reflects its rapid disk formation and settling, but increasing age uncertainties shift this feature to younger ages.
Ultimately, the impact of age uncertainties varies between galaxies, changing our inference about the ISM settling history and dynamical heating of old stars (weakly for m12m, strongly for Romeo), the merger history (m12f), and the disk formation/settling time (Louise).}
\label{fig:cases}
\end{figure*}

Figure~\ref{fig:cases} shows the total stellar velocity dispersion, $\sigma_{\rm 3D}$, versus stellar age, $\tau$ in four of our galaxies: m12m (top left), m12f (top right), Romeo (bottom left), and Louise (bottom right). 
We selected these galaxies to bracket the mass range of the observed systems that we compare against in Section~\ref{sec:P2} -- namely the MW, M31, and M33 -- and to serve as useful analogs: Louise is comparable in mass to M33, Romeo resembles the MW, and m12m is a comparable stellar mass to M31.
We also include m12f to illustrate a merger-driven spike in its $\sigma(\tau)$ and to demonstrate how such features are affected by age uncertainties.

All galaxies show a characteristic rise in $\sigma(\tau)$ with age, the result of disk settling and post-formation dynamical heating, as \citep{McCluskey} explored for these galaxies.
The youngest stars exhibit the most rapid increase, because heating is most efficient for dynamically cold populations \citep{bt-08}.
However, the exact shape of $\sigma(\tau)$ reflects the unique history of each galaxy and, because of this, age uncertainties impart different effects on each galaxy.

With a stellar mass of $10^{11} \Msun$, m12m is our most massive galaxy and the closest to M31's stellar mass of $10.3 \pm 2.3 \times 10^{10} \Msun$ \citep{Sick15}.
In m12m, $\sigma$ rises steadily with age between 1 and 7 Gyr, which in this case primarily reflects the continued settling of the star-forming ISM \citep{McCluskey}.
For stars younger than 6 Gyr, the impact of age uncertainties is weak, while for older ages, age uncertainties reduce $\sigma$ and flatten its age dependence.

m12f highlights the dramatic impact that a gas-rich merger can have and the equally dramatic impact that age uncertainties have on the observability of this feature.
m12f underwent a merger (of stellar mass ratio $\approx$1:13) $\approx 6.9 \Gyr$ ago, after it entered its Early-Disk Era, but before it settled into a thin disk $\approx 4.6 \Gyr$ ago.
This merger led to a significant spike in the \textit{present-day} $\sigma$ of stars formed around that time.
When measured with perfect age precision, this spike is dramatic, and it would provide a direct way to age-date the merger.
However, even 10\% age uncertainties reduce this spike to a gentle bump, and 20\% age uncertainties wash it out entirely.

Not all mergers drive such spikes in $\sigma$ today.
m12f underwent another merger $\approx 500 \Myr$ ago with an LMC-mass satellite.
This merger was by no means negligible: for example, \cite{Necib19} shows that this merger accounts for 53\% of the accreted stellar mass in the solar circle, while \cite{Yu2021} show that this merger triggered a significant burst of star formation.
Nevertheless, $\sigma(\tau)$ remains comparatively smooth, even in the absence of age uncertainties.
Thus, merger signatures in $\sigma(\tau)$ depend on the timing and orbit of the merger.
The fact that mergers that coincide with (and perhaps facilitate) disk formation can lead to spikes in $\sigma(\tau)$ is particularly relevant for the MW, given that its disk formation was potentially coincident with the GSE merger \citep[for example][]{Conroy22}.
Correctly characterizing such merger signatures require age uncertainties of $\lesssim 10\%$, with the earlier timing of GSE ($\approx 10 \Gyr$ ago) potentially necessitating even tighter constraints within an already enigmatic age regime.

Romeo is our earliest-forming galaxy, both in terms of stellar mass assembly and the onset of disk formation, which occurred $\approx 11 \Gyr$ ago \citep{McCluskey}.
Thus, in many ways, it provides our best MW analog, in terms of an early-forming disk \citep{Belokurov22, Xiang22}, being in an LG-like environment, and having a stellar mass of $5.9 \times 10^{10} \Msun$.
In the absence of age uncertainties, Romeo's $\sigma_{\rm 3D}$ increases steadily with age, and with a sharper rise $\approx 8$ and 12 Gyr ago, which is the steepest relation for old stars among our sample.
Thus, for Romeo, \textit{fractional} age uncertainties significantly reduce $\sigma$ in older stars by up to $50 - 75 \kms$. Given that Romeo is our best dynamical analog to the MW, this suggests particular caution in interpreting $\sigma$ of old stars in the MW.

Louise is the lowest-mass galaxy in our sample, with $M_{\rm star} \approx 2.3 \times 10^{10} \Msun$, and is thus closest to approaching the even lower mass of M33 -- $4.8 \times 10^{9} \Msun$ \citep{Corbelli14}. For Louise, $\sigma$ rises gradually with age up to $6 \Gyr$, but $\sigma$ increases rapidly between $\approx 6$ and 9 Gyr ago, reflecting the rapid settling of its disk.
However, increasing age uncertainties shift this rapid rise to occur more recently, by up to 2 Gyr.
In contrast, at ages $\gtrsim 9 \Gyr$, the flatness of $\sigma(\tau)$ leads to only minor effects for age uncertainties.

Ultimately, the impact of age uncertainties varies between galaxies and depends on a galaxy's formation history.
Age uncertainties can change our inference about the ISM settling history and dynamical heating of old stars (weakly for m12m, strongly for Romeo), the merger history (m12f), and the disk formation/settling time (Louise).

\subsubsection{Trends across all simulations}
\label{subsubsec:sample wide} 

\begin{figure}
\centering
\includegraphics[width = \columnwidth]{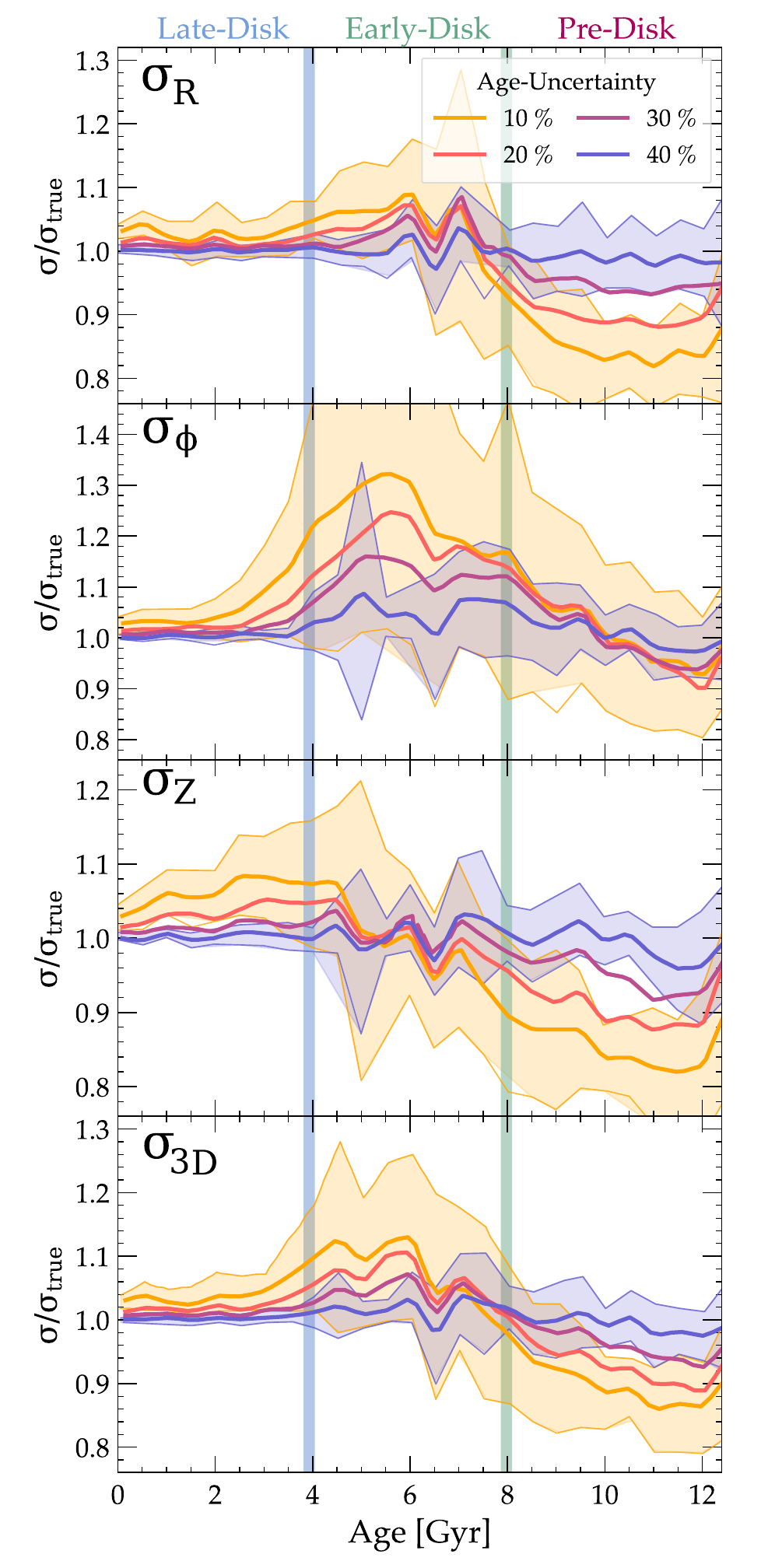}
\caption{
\textbf{The impact of stellar age uncertainties on the measured velocity dispersion.}
The ratio of $\sigma$ using different age uncertainties to its value without any age uncertainty, for $\sigma_{\rm 3D}$, $\sigma_{R}$, $\sigma_{\phi}$, and $\sigma_{Z}$.
Lines show the average across our 11 galaxies, and shaded regions show their corresponding 16-84th percentile range (for 10\% and 40\% age uncertainties).
The shaded vertical bars indicate when these galaxies transitioned, on average, from the Pre-Disk to the Early-Disk Era ($\approx 8 \Gyr$ ago) and from the Early-Disk to the Late-Disk Era ($\approx 4 \Gyr$ ago).
Overall, age uncertainties have only a minimal impact on $\sigma$ of Late-Disk stars, with 40\% uncertainties leading to $\lesssim 10\%$ increases.
For Early-Disk stars, $\sigma_{R}$ and $\sigma_{Z}$ continue to be only minorly impacted, while $\sigma_{\phi}$ noticeably increases -- on average, 20 and 40\% age uncertainties cause $\sigma_{\phi}$ of 6 Gyr old stars to be overestimated by 15 and 30\%, reflecting the (usually) rapid growth of the disk during this period.
For Late-Disk stars, age uncertainties lead to lower $\sigma$ on average, although this effect is within 10\% for uncertainties $\lesssim 30\%$.
Ultimately, age uncertainties have only a minor impact on the \textit{value} of $\sigma$ at a given age: on average, $\sigma_{\rm 3D}$ remains within 20\% of its true value, even for age uncertainties of 40\%.
}
\label{fig:unc}
\end{figure}

Figure~\ref{fig:unc} shows how uncertainties in stellar age affect $\sigma(\tau)$ on average across 11 FIRE-2 galaxies.
Because the impact varies by velocity component, we show radial, tangential, and vertical $\sigma$ separately.
For each galaxy, we compute $\sigma(\tau)$ with and without age uncertainties in individual apertures, normalize by the median value for that galaxy, and then average across the sample.
This normalization allows for more meaningful comparisons across galaxies with varying stellar masses, star formation histories, and disk structures.
Shaded regions denote the $\rm 16^{th}$–$\rm 84^{th}$ percentile range, illustrating the degree of galaxy-to-galaxy scatter, which is particularly prominent for young stars.
For context, shaded bars along the top of the figure indicate the average lookback times when these galaxies transitioned between the Pre-Disk, Early-Disk, and Late-Disk Eras (see Section~\ref{subsec:methods_age}).

Figure~\ref{fig:unc} shows that the impact of age uncertainties varies depending on when the stars formed relative to the disk eras.
In general, age uncertainties \textit{increase} $\sigma$ for stars formed in the Late-Disk and Early-Disk Eras and decrease $\sigma$ for stars formed in the Pre-Disk Era.
For Late-Disk stars, age uncertainties of $20\%$ increase $\sigma$ by $\lesssim 2\%$, while uncertainties of $40\%$ increase $\sigma$ by $\lesssim 10\%$.
However, age uncertainties increase $\sigma$ for Early-Disk stars more significantly. In particular, $\sigma_{\phi}$ shows the strongest increase with age uncertainty, up to $30\%$, during the Early-Disk Era. In contrast, $\sigma_{R}$ increases by up to $\approx 10\%$, and $\sigma_{Z}$ changes by $\lesssim 5\%$. 
The disproportionate impact on $\sigma_{\phi}$ results from the fact that $v_{\rm \phi}$ shows the strongest time evolution (both across all components and ages) throughout the Early-Disk Era due to the (generally rapid) onset of disk formation. Although the median $v_{R}$ and $v_{Z}$ of $\sim$4 and 8 Gyr old stars are usually broadly similar at $\sim$0 km/s, the median $v_{\phi}$ can differ by up to $\sim$200 km/s. Thus, the faulty inclusion of older, non-disky stars artificially increases $\sigma_{\phi}$. 
Lastly, age uncertainties reduce $\sigma$ by up to $10 - 20\%$ for Pre-Disk stars, with $\sigma_{\phi}$ exhibiting the weakest decrease.

Overall, age uncertainties most significantly impact $\sigma_{\phi}$, inflating its value by up to $30\%$ at intermediate ages ($4 - 8 \Gyr$ old). In comparison, the effects on $\sigma_{R}$ and $\sigma_{Z}$ are generally weaker. Ultimately, the effect of age uncertainties on the total $\sigma_{\rm 3D}$ is generally $\lesssim 15\%$ at any given age.
However, we repeat that these are the \textit{average} shifts in $\sigma$ at fixed age across our galaxy sample. Within individual galaxies, the effect of age uncertainties can be more dramatic, shifting the inferred ages of certain kinematic features (as shown in Figure~\ref{fig:cases}).

\subsection{Aperture Size}
\label{subsec:ap_size} 

Next, we study how measurements of $\sigma(\tau)$ depend on physical scale.
The long-known size-linewidth relation, that $\sigma_{\rm GMC} \propto l^{0.5}$ \citep{Larson81}, indicates that $\sigma$ of molecular gas increases with the size of the region, often as a power law \citep{Heyer15, Rice16, Rosolowsky21, Leroy25}.
%though with possible systematic variation with galactic environment \citep{Rosolowsky21, Chevance23, Schinnerer24, Leroy25}.
This arises because the star-forming ISM is supersonically turbulent, and GMCs reflect this hierarchical turbulent structure.
The FIRE-2 simulations show this behavior in the ISM, with various works confirming that star-forming gas and GMCs in the simulations follow the observed size-linewidth relation \citep{hopkins18, hung19, Guszejnov20}.

Stars form from dense gas and thus inherit the clustered distribution of their natal ISM.
In turn, $\sigma$ of newly-formed stars is correspondingly scale-dependent \citep{LadaLada03, Krumholz19, Maschmann24}.
However, this inherited scale dependence rapidly evolves and is erased on timescales of $\approx 100 \Myr$, as stellar orbits mix and evolve \citep{Grasha19, Peltonen23}.

\cite{McCluskey} showed in these FIRE-2 simulations that $\sigma$ of young stars (and cold gas) increased by $\approx 2 \times$ when measured across an annulus versus when measured in small ($150 \pc$) apertures, in part because larger apertures incorporate later-scale disk dynamical structures such as spiral arms, bars, and warps.
This aperture dependence provides an important context for interpreting and comparing observational work.
Historically, many studies of the MW focused on stars within a few 10's of pc of the Sun, while more recent surveys extended to a few kpc.
In M31 and M33, where observations span considerable swaths of the disk, observations use smoothing techniques to measure $\sigma$ within apertures of radius $\approx 0.5 \kpc$.
Furthermore, high-resolution studies of nearby galaxies, such as PHANGS-MUSE, derive $\sigma$ from integrated spectra of $100 – 200 \pc$ spaxels.
%With new observations spanning multiple spatial scales, we assess how measurement choices influence our understanding of galactic structure and evolution.

\begin{figure}
\centering
\includegraphics[width = \columnwidth]{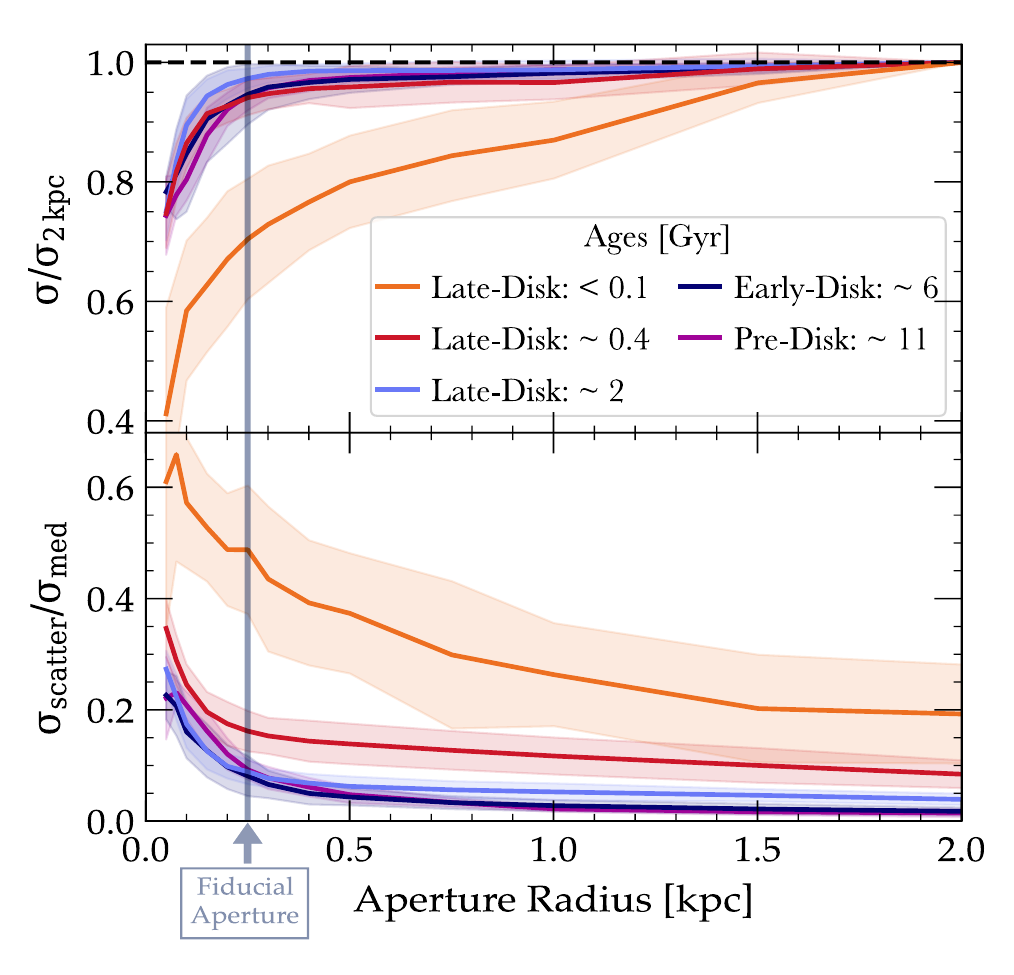}
\vspace{-4 mm}
\caption{
\textbf{Effect of aperture radius, $r_{\rm ap}$, on the measured 3D velocity dispersion, $\sigma_{\rm 3D}$.}
We measure $\sigma_{\rm 3D}$ in cylindrical apertures of radius $r_{\rm ap}$ and $|Z| < 3 \kpc$ across an annulus centered at $R = 8 \kpc$. 
We average across 11 FIRE-2 galaxies, and the shaded region shows the galaxy-to-galaxy standard deviation.
\textbf{Top}: The median of $\sigma_{\rm 3D}$ versus $r_{\rm ap}$ for stars of different ages, normalized by the value for a large aperture of $r_{\rm ap} = 2 \kpc$.
The dashed horizontal lines indicates 100\% of the dispersion value for $r_{\rm ap} = 2 \kpc$.
$\sigma_{\rm 3D}$ is nearly independent of the aperture size at $r_{\rm ap} \gtrsim 0.25 \kpc$ for all but the youngest stars (age $\lesssim 100 \Myr$).
For these stars, $r_{\rm ap} < 0.25 \kpc$ leads to $\sigma$ that are $\approx 50\%$ of those measured with $r_{\rm ap} = 2 \kpc$.
\textbf{Bottom}: The fractional aperture-to-aperture scatter in $\sigma$ around the annulus, normalized by the median $\sigma$.
$\sigma_{\rm scatter} / \sigma_{\rm med}$ increases with smaller aperture radius, especially at $r_{\rm ap} \lesssim 0.25 \kpc$.
The youngest stars have the largest relative contribution from $\sigma_{\rm scatter}$.
For $r_{\rm ap} = 250 \pc$, these $\sigma$ vary by $\approx 50\%$ across the annulus.
Older stars exhibit small scatter, indicating these populations are well mixed across the annulus.
}
\label{fig:scatter}
\end{figure}

To quantify the impact of the aperture radius, Figure~\ref{fig:scatter} (top) shows the ratio of $\sigma_{\rm 3D}$ measured in a cylindrical aperture of radius $r_{\rm ap}$ to its value when measured within $r_{\rm ap} = 2 \kpc$, for five representative age ranges.
In general, the impact of aperture radius is weak: for all but the youngest stars, apertures of 50 pc yield $\sigma \approx 80\%$ as large as those in 2 kpc apertures, while apertures of 150 pc and 250 pc yield $\sigma$ that are 90\% and 95\% as large, respectively.
Ultimately, aperture radii larger than $\approx 500 \pc$ do not matter.
These results indicate that stars older than 100 Myr are well phase-mixed in the FIRE-2 simulations.
Furthermore, we interpret the sharp downturn below 200 pc for stars older than $\approx 200 \Myr$ as reflecting the numerical limits from resolution in the FIRE-2 simulations, so we use $250 \pc$ as our fiducial aperture radius throughout.

That said, the youngest stars (ages $\lesssim 100 \Myr$) show stronger aperture dependence.
Aperture radii of 50 and 250 pc yield median $\sigma \approx 45\%$ and 70\% of their value at 2 kpc, and they only reach 90\% at $1.25 \kpc$.
This gradual increase of $\sigma$ with aperture radius, out to scales $\gtrsim 1 \kpc$, highlights the \textit{physical effects} of clustered star formation and gas turbulence. However, the absence of this scale dependence for stars $100 – 600 \Myr$ old suggests that newly formed stars rapidly diffuse through phase space, with clusters dispersing on the order of a dynamical time.
Thus, one should exercise caution in comparing results that measure $\sigma$ of young stars across different spatial extents.

%We find that stars have local present-day velocity dispersions that are significantly lower than their disk-wide counterparts. For example, stars younger than 100 Myr at R=8 kpc have local dispersions that equal $\approx$ 57, 55, 77, and 63\% of their disk-wide $\sigma_{R}$, $\sigma_{\phi}$, $\sigma_{Z}$, and $\sigma_{\rm tot}$, respectively, corresponding to respective differences of $\sim$ 14, 11, 3, and 16 km/s. Therefore, $\approx$ 37\% of these stars' $\sigma_{\rm tot}$ arises from spatial variations within the disk. 

Figure~\ref{fig:scatter} (bottom) quantifies how aperture size affects the aperture-to-aperture \textit{scatter} in the measured $\sigma_{\rm 3D}$ at fixed $R$.
It shows the \textit{fractional scatter} via the ratios of the scatter (the 68th percentile width) of $\sigma_{\rm 3D}$, relative to $\sigma_{\rm med}$, around an annulus for a given stellar population.
For all but the youngest stars, $\sigma_{\rm scatter} / \sigma_{\rm med}$ is remarkably scale-independent at $r_{\rm ap} \gtrsim 250 \pc$, being $\approx 10\%$, 5\%, and 2\% of the typical $\sigma_{\rm 3D}$ at ages $400 \Myr$, $2 \Gyr$, and $\gtrsim 6 \Gyr$.   
%For the middle and oldest age groups, small apertures yield scatters that are about 20\% of the typical dispersion, which drop to less than 5\% for aperture radii greater than 0.4 kpc.
As above, the aperture-to-aperture scatter is most significant for young stars: $\sigma_{\rm 3D}$ within $r_{\rm ap} = 250 \pc$ varies by $\approx 50\%$ across the annulus, and remains as high as 20\% even for $r_{\rm ap} = 2 \kpc$.

These results help to contextualize how the stellar $\sigma$ varies within different apertures and ''solar neighborhoods'' of a galaxy.
For example, as we will show in Section~\ref{subsec:comparing observations}, the MW exhibits significantly colder kinematics than other galaxies.
Individual 250 pc regions can have $\sigma$ for \textit{young} stars that are up to 40\% lower than the galaxy's average at a given $R$, which may be a limiting systematic in interpreting the MW.
That said, such aperture-to-aperture variations do \textit{not} provide a key systematic uncertainty for \textit{older} stars.

\subsection{Galactocentric Radius}
\label{subsec:rad_selection} 

Observations show that $\sigma$ decreases with galactocentric radius, $R$, in both external disk galaxies \citep[for example][]{Bottema93, Kregel05} and the MW \citep[for example][]{Dehnen06, Bovy12}.
A classical explanation for this decrease follows from $\Sigma(R) \propto \sigma_{Z}^{2} / h_{Z}$ for a disk in equilibrium, where $\Sigma(R)$ is the surface density and $h_{Z}$ is the disk scale height.
To the extent that $h_{Z}(R)$ does not change significantly with radius, while $\Sigma(R)$ falls off rapidly with $R$ (often as an exponential), this implies that $\sigma_{Z}$ should decline with $R$.
%This structure broadly reflects the ``inside-out'' formation of galactic disks: star formation advances from a galaxy's central regions to larger and larger radii over time due to the continued accretion of high angular momentum gas.

\begin{figure}
\centering
\includegraphics[width = \columnwidth]{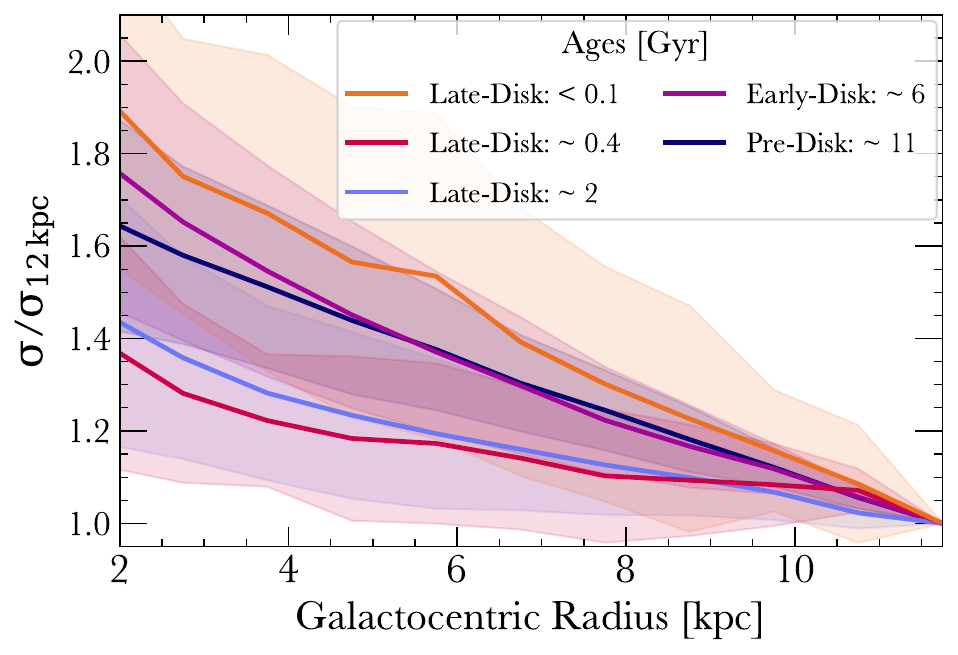}
\caption{
\textbf{Effect of galaxy radial selection on the measured 3D velocity dispersion.}
$\sigma_{\rm 3D}$, normalized to its value at $R = 12 \kpc$, for stars of different ages. 
$\sigma_{\rm 3D}$ increases towards the inner galaxy, with the youngest stars (age $\lesssim 100 \Myr$) having the strongest radial dependence, followed by Early- and Pre-Disk stars, and finally the older Late-Disk populations, indicating that though the star-forming ISM is more turbulent at small radii, subsequent radial mixing of young stars mitigates this dependence.
}
\label{fig:rad_ap}
\end{figure}

Figure~\ref{fig:rad_ap} shows $\sigma_{\rm 3D}$, measured in cylindrical apertures of $r_{\rm ap} = 250 \pc$ and vertical extent $|Z| < 3 \kpc$, versus $R$ for stars of different ages.
We limit our analysis to $R = 2 - 12 \kpc$, neglecting the innermost region where the tangled contributions from the bulge, bar, and inner halo complicate interpreting the disk component; as well as the outer disk, where the number of older star particles is limited.
$\sigma$ increases with age at all $R$, and this is the primary dependence.
Thus, to isolate trends with $R$, in Figure~\ref{fig:rad_ap} we normalize the $\sigma$ at each age to its value at $R = 12 \kpc$.
We have verified that our results are not sensitive to this normalization, as even in the (potentially warped) outer disk, the dispersion depends just on the relative motion of stars in that patch.%, meaning the dispersion of a ``warped'' patch is not inflated just by the virtue of being a warp. 

At all ages, $\sigma$ decreases with $R$: stars in the inner galaxy have hotter kinematics.
Such behavior is expected, given that previous work confirmed that the gas mass, stellar mass, and star formation surface densities in these simulations decrease with $R$ \citep{Gurvich20, Orr20}.
The strength of this radial dependence exhibits a complicated relationship with age: the youngest stars (age $< 100 \Myr$) have the strongest radial dependence, while stars of ages $\approx 6 \Gyr$, 11 Gyr, 400 Myr, and 2 Gyr show progressively weaker radial gradients.

The youngest stars (age$<$100 Myr) show the strongest radial dependence: young stars in the inner galaxy have nearly twice the $\sigma$ as their cohorts in the outer galaxy. This strong radial dependence is largely – though not exclusively – inherited from the star-forming ISM.  
In these FIRE-2 simulations, the gas surface density, $\Sigma_{\rm gas}$, generally decreases by an order of magnitude between $R = 2 - 12 \kpc$ \citep{Graf24b}.
This outward radial decline in $\sigma$ also results from radially-dependent post-formation heating, and post-formation dynamical heating may be stronger in the inner galaxy \citep{Schronich09, Aumer16}.

These trends also reflect the radial redistribution of these populations. 
Stars do not necessarily remain at the $R$ where they formed, but, via various dynamical processes, can redistribute across multiple kpc \citep[for example][]{Sellwood_Binney02, Roskar11, Vera-Ciro14, Daniel18, el-badry18, Frankel20}.
In contrast to dynamical heating, post-formation radial redistribution has a negligible effect on stars younger than $\approx 100 \Myr$.
This is not necessarily the case for older stellar populations. Indeed, 400 Myr and 2 Gyr old stars show the \textit{weakest} radial dependence, despite forming with a radial dependence similar to that of 100 Myr old stars, due to radial redistribution.
While stars from the cold outer disk exhibit a net inward redistribution, stars from the hot inner disk exhibit a net outward distribution (Bellardini et al. in prep.).
In turn, this mixing subsequently weakens the strong radial dependence inherited at formation.

Stars 6 and 11 Gyr old exhibit a stronger radial dependence than those 400 Myr and 2 Gyr old.
This might seem to contradict our earlier claim, given that these stars had more time to redistribute and weaken any initial radial dependence.
However, when these stars formed, these galaxies were more radially compact, vertically thick, and highly turbulent \citep[for example][]{El-Badry18-anc, Horta24}.
Over time, these galaxies grew and formed disks, with subsequent generations of stars forming in increasingly thin and radially extended disk \citep{Yu22, Carrillo23, Bellardini22}.
Thus, while Late-Disk stars formed throughout this radial range ($R = 2 - 12 \kpc$), older stars predominantly formed in the (now) inner galaxy.

The stronger radial dependence exhibited by stars that formed in the Early-Disk and Pre-Disk Eras compared with those that formed in the Late-Disk Era (save the youngest population) agrees with observations.
For example, \cite{Aniyan18} shows that in the nearby spiral galaxy, NGC 628, stars older than 3 Gyr exhibit a $\approx 50\%$ stronger radial dependence than their younger counterparts.

\subsection{Galaxy Inclination Angle}
\label{subsec:inc}

The MW is the only galaxy where we can directly measure all three velocity components; in external galaxies, observations are typically limited to the 1D line-of-sight velocity, $v_{\rm LOS}$. Because $v_{\rm LOS}$ is a projection of the intrinsic 3D velocity, its value depends on inclination, with different viewing angles amplifying or suppressing individual velocity components. The inclination dependence of $\sigma_{\rm LOS}$ complicates our ability to directly compare galaxies, and may obscure (or erroneously suggest) intrinsic differences in their kinematics.

For example, \citet{McCluskey} examined the evolution of these components in these FIRE-2 simulations and found that $\sigma_{R} > \sigma_{\phi} > \sigma_{Z}$ at all ages, both at formation and today.
At birth, the radial component accounted for about half of the total energy in $\sigma$, with the azimuthal and vertical components contributing $\approx 32\%$ and $\approx 17\%$, respectively.
Over time, post-formation heating amplified these differences, with young stars experiencing stronger radial heating, likely due to bar and spiral structure, while vertical heating remained more constant.
As a result, while $\sigma_{Z}$ shows the weakest absolute increase with age, it exhibits the strongest fractional evolution.

Rather than deriving full corrections for inclination effects -- where dust extinction and beam smearing introduce additional complexities -- we isolate the geometric influence of inclination, specifically how $\sigma_{\rm LOS}$ changes due to the shifting contributions of $\sigma_{\rm R}$, $\sigma_{\rm \phi}$, and $\sigma_{\rm Z}$ (see \cite{Ejdetjarn22} for discussion on additional inclination effects on gas dispersions).
Most importantly, we show how this effect varies with stellar age, providing insight for comparing $\sigma(\tau)$ across different galaxies.\footnote{Our spatial selection does not account for inclination-induced projection effects: we define cylindrical apertures consistently across all inclinations, ensuring that each aperture always contains the same stars.
This effectively selects apertures as if viewing the galaxy face-on, where a single line of sight corresponds to an aperture along the annulus.
In reality, at higher inclinations, each line of sight cuts through a cylindrical stripe of the disk, so an aperture includes stars from a range of galactocentric radii and azimuthal positions.}

\begin{figure}
\centering
\includegraphics[width = \columnwidth]{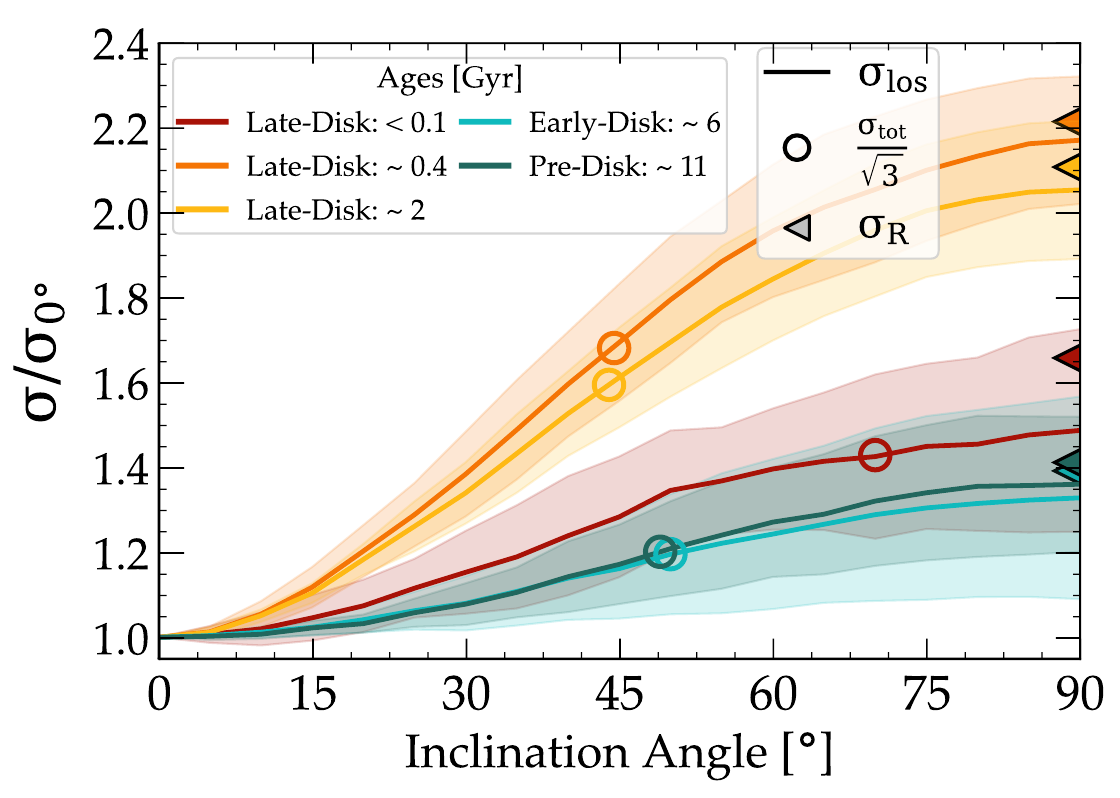}
\caption{
\textbf{The impact of galaxy inclination angle on the line-of-sight velocity dispersion.}
$\sigma_{\rm LOS}$, normalized to its value for a face-on galaxy (0$^\circ$), versus inclination angle for stars of various ages.
Circles show where $\sigma_{\rm LOS} = \sigma_{\rm 3D} / \sqrt{3}$.
For reference, triangles at right mark $\sigma_{R}$, which is the largest component, at each age.
For all ages, $\sigma_{\rm LOS}$ increases with inclination, equaling $\sigma_Z$ at 0$^\circ$ but remaining slightly subdominant to $\sigma_R$ at 90$^\circ$.
Older stars generally show weaker dependence on inclination, reflecting their less circular orbits, while 0.4 and 2 Gyr old stars show the strongest dependence, such that their edge-on $\sigma$ are $\gtrsim 2 \times$ of their face-on values.
Interestingly, the inclination dependence of the youngest stars (age $\lesssim 100 \Myr$) is intermediate to the older Late-Disk and Early/Pre-Disk populations, indicating that the star-forming ISM is more isotropic and that early heating is primarily in-plane.
}
\label{fig:inc}
\end{figure}

Figure~\ref{fig:inc} shows $\sigma_{\rm LOS}$, normalized to its value for a face-on projection, versus inclination for stars of different ages. 
At each age, we mark the value of $\sigma_{R}$ with a triangle on the right, because $\sigma_{R}$ is the largest, so it sets an upper limit on $\sigma_{\rm LOS}$.
We also indicate the inclination where $\sigma_{\rm LOS} = \sigma_{\rm 3D} / \sqrt{3}$ with circles, in part to aid in our later comparisons to the MW, for which we assume isotropy in comparing with other observed galaxies.

At all ages, $\sigma_{\rm LOS}$ increases with inclination. This is unsurprising: in face-on galaxies, $\sigma_{\rm LOS}$ reflects only vertical motions, while at edge-on inclinations, it is entirely independent of $\sigma_{Z}$.
Because $\sigma_{Z}$ is smaller than the in-plane $\sigma$ in disk galaxies, higher inclinations amplify the contribution of the dominant in-plane components to $\sigma_{\rm LOS}$.

Inclination effects depend on stellar age: $\sigma_{\rm LOS}$ of 400 Myr and 2 Gyr old stars shows the strongest dependence on inclination, with edge-on $\sigma_{\rm LOS}$ exceeding face-on $\sigma_{\rm LOS}$ by a factor of $\approx 2$.
By comparison, young stars (age $< 100 \Myr$) show a weaker dependence on inclination.
The stronger inclination dependence after just a few 100 Myr reflects the nature of early dynamical heating of these stars: rapid and strongly in-plane.
$\sigma_{\rm 3D}$ increases from $30 \kms$ at $100 \Myr$ to $75 \kms$ at $1 \Gyr$, with $\sigma_R$ accounting for 60\% of this increase, $\sigma_\phi$ by 30\%, and $\sigma_Z$ just under 10\% (see Figure~\ref{fig:unscaled}).
This radially-biased dynamical heating becomes weaker over time, while vertical heating persists, as Early-Disk and Late-Disk stars exhibit weaker inclination dependencies and thus more isotropic $\sigma$ (see Section~\ref{subsubsec:fire mw}).

Although many methods for inclination correction exist \citep[for example][]{Westfall19}, they typically assume a homogeneous stellar population, applying the same correction across all stars regardless of age.
However, Figure~\ref{fig:inc} shows that the inclination affects stellar populations of different ages in distinct ways. 
A one-size-fits-all correction can misrepresent the intrinsic kinematics by overestimating $\sigma$ of younger stars (400 Myr to 2 Gyr) relative to both the youngest and older stars.
Overall, one must apply age-dependent corrections before interpreting trends such as dynamical heating or other evolutionary processes.

\section{Observational Comparisons}
\label{sec:P2}

Thus far, we used FIRE-2 simulations to investigate how age uncertainties, aperture size, galactocentric radius, and galaxy inclination influence measurements of stellar velocity dispersion.
These results provide us with a framework for combining and comparing disparate observations of $\sigma(\tau)$ in observed galaxies.
We now compile observations from the literature, including the MW, M31, M33, and PHANGS galaxies, and we compare $\sigma(\tau)$ in these galaxies.
We then use these observations to benchmark how well the FIRE-2 simulations model $\sigma(\tau)$.

\subsection{M31, M33, and PHANGS}
\label{subsec:M31}

In this section, we outline three works that measured $\sigma_{tau}$ in disk galaxies other than the MW: the studies of M31 by \citet{Dorman2015} (hereafter \citetalias{Dorman2015}), M33 by \citet{Quirk22} (hereafter \citetalias{Quirk22}), and 19 PHANGS-MUSE galaxies by\citet{Pessa23} (hereafter \citetalias{Pessa23}).
Here we provide a broad overview of the observations and briefly discuss certain key details of each work's method, and we discuss these observations in greater detail in Appendix~\ref{a:external_obs}.
Table~\ref{table:table_obs} lists these galaxies' stellar masses, circular velocities, inclinations, and the typical aperture radius used by each study.

\begin{table*}
\caption{
properties of nearby disk galaxies with measured stellar velocity dispersion versus age
}
\setlength\extrarowheight{2.3pt}
\centering
%\hspace{-20 mm}
\begin{tabular}{|c|c|c|c|c|c|}
\hline
Galaxy & $M_{\star}$ &  $v_{\rm \phi,0} $ & $r_{\rm ap}$ & Inclination & Reference \\
& $\mathrm{[10^{10} M_\odot]}$ & [km/s] & [pc] & [\degree] & \\
\hline
Milky Way & 5 $\pm 1^{\rm a}$ & $238^{\rm a}$ & 250 & - & See Table~\ref{table:table_mw} \\
M31 & $10.3 \pm 2.3^{\rm b}$ & 260 & 760 & 77 & \cite{Dorman2015} \\
M33 & $0.48 \pm 0.06^{\rm c}$ & 93 & 420 & 54 & \cite{Quirk22} \\
PHANGS median (range) & 3.63 (0.25 - 9.3) & $182 (137 - 230)^{\rm d}$ & 100 & 38.5 (8.9 - 57.3) & \cite{Pessa23} \\
\hline
\end{tabular}
\tablecomments{
Values without superscripts are from the listed reference, while superscripted values are from: a:~\cite{bhg-16}, b:~\cite{Sick15}, c:~\cite{Corbelli14}, and d:~\cite{Lang20}. $v_{\phi,0}$ refers to the maximum circular velocity of the gaseous or young stellar disk (see text for more details). 0$\degree$ inclination is face-on, 90$\degree$ is edge-on.
}
\label{table:table_obs}
\end{table*}

M31's proximity ($785 \kpc$), nearly edge-on inclination ($77^\circ$), and low foreground extinction \citep{Schlafly11} provide a relatively clear view of its entire disk and halo, while also allowing observations of many individually resolved stars.
Like the MW, M31 hosts a central bar and an extended stellar disk. However, multiple lines of evidence indicate that our massive neighbor had a more extended and active accretion history. For example, M31's stellar halo is $\approx 10 \times$ more metal-rich and $\approx 10 \times$ more massive than the MW's, containing significant tidal debris and a metal-rich component consistent with belonging to a ``kicked-up" disk population \citep{Dorman13, Escala23}. Similarly, M31's disk appears thicker than the MW's, with RGB stars having a scale height of $\approx 770 \pc$ \citep{Dalcanton23}. Simulations indicate that these properties are consistent with M31 experiencing a major ($\approx 4:1$) merger $\approx 2.5 \Gyr$ ago \citep{D'Souza18, Hammer18}, which is likely related to M31's contemporaneous global burst of star formation \citep[for example][]{Williams15}.

Much of this knowledge of M31 results from large-scale surveys of resolved stellar populations throughout M31's disk and halo. Two notable surveys of M31's disk are the Panchromatic Hubble Andromeda Treasury (PHAT) survey \citep{Dalcanton12, Williams14}, a Hubble Space Telescope (HST) MultiCycle Treasury program that provided six-filter photometry for 138 million stars, and the Spectroscopic and Panchromatic Landscape of Andromeda’s Stellar Halo (SPLASH) survey \citep{Guhathakurta06, Gilbert12, Dorman12}, which obtained Keck/DEIMOS spectra for $\approx 10,000$ stars.
These two surveys allowed for the first measurements of $\sigma(\tau)$ in a disk galaxy other than the MW.

M31's largest satellite, M33, allows us to study a disk at lower masses. Lying $859 \kpc$ from the MW, M33 offers a holistic view and resolved stellar populations. With a stellar mass of $3 \times 10^{10} \Msun$, M33 provides a crucial perspective on the formation of galactic disks, as this is precisely the ``mass of disk formation'', that is, the mass where the observed (stellar mass) Tully-Fisher relation transitions \citep{Simons15}.
HST provided six-filter photometry for millions of disk stars, here for $\approx 22$ million stars via the Panchromatic Hubble Andromeda Treasury: Triangulum Extended Region (PHATTER) survey \citep{Williams21}, while complementary spectroscopic surveys provide line-of-sight velocities for a small fraction of the photometric sample. For example, the TRiangulum EXtended Survey (TREX) survey \citep[][hereafter \citetalias{Quirk22}]{Quirk22} measured spectra for $\approx 7000$ stars across the disk of M33.
Importantly, these two surveys allowed \citetalias{Quirk22} to measure $\sigma(\tau)$ in M33 using $\approx 3000$ stars.

Within the LG, we can resolve \textit{individual} stars, but the number of disk galaxies is limited.
Fortunately, integral-field spectroscopy (IFS) of nearby galaxies offers a compelling new window, with IFS instruments like MUSE on the VLT providing spatially resolved spectra of stellar populations on $\approx 100 \pc$ scales.
Recently, \citetalias{Pessa23} used MUSE to measure $\sigma(\tau)$ in the 19 star-forming galaxies comprising the PHANGS-MUSE sample.
These galaxies lie within $\lesssim 20 \Mpc$ of the MW, have low to moderate inclinations ($< 60^{\circ}$), and have stellar masses of $\log(M_\star / M_\odot) = 9.4 - 11.0$, with a median of $\log(M_\star / M_\odot) = 10.5$.
Although \citetalias{Pessa23} measured $\sigma(\tau)$ for all 19 of these galaxies, we only include results from the 16 that also have reported rotational velocity measurements from \citet{Lang20}.
An important caveat for interpreting the results of this work is that these $\sigma$ from MUSE are likely overestimated at small intrinsic values (and thus at young ages), as we discuss in Section~\ref{subsec:comparing observations}.

\subsection{The Milky Way}
\label{subsec:mw observations}

We now present a compilation from the literature of observational measurements of $\sigma(\tau)$ of the MW.
We restrict our compilation to results published after the release of \textit{Gaia} DR1, with the exception of three commonly cited studies based on the (pioneering and pivotal) Geneva-Cophenhagen Survey (GCS).
Table~\ref{table:table_mw} summarizes the key properties of the observational datasets we include.
As for the above galaxies, we provide additional details in Appendix~\ref{a:mw}.

Figure~\ref{fig:mw obs} shows the radial, azimuthal, vertical, and 3D velocity dispersions in the MW from 13 analyses. While some works present all 3 velocity components, many present only one or two.
Consequently, each panel of Figure~\ref{fig:mw obs} includes different subsets of observations. Specifically, while $\sigma_{Z}$ contains data from all 13 works, $\sigma_{R}$, $\sigma_{\phi}$, and $\sigma_{\rm 3D}$ include data from 9, 7, and 8 works, respectively.
Secondly, 5 of these works split their sample into separate high- and low-$\alpha$ populations and did not present $\sigma$ for their combined sample.
For these works, we show the low-$\alpha$ data as filled markers and the high-$\alpha$ data as unfilled.

In addition to showing $\sigma$ reported by each of these works (scatter points), Figure~\ref{fig:mw obs} also shows the mean of the sample (solid lines) and its maximum-minimum range (shaded regions).
To standardize the age-binning of these different observations, we used a 1-D linear interpolation method to fit each work's data into 250 Myr-wide age bins spanning $0.25 - 13.25 \Gyr$. 
For works that distinguished between high- and low-$\alpha$ stars, we treated the data of the two populations separately.
We then found the mean/minimum/maximum $\sigma$ of our interpolated sample in each age bin, excluding data for high-$\alpha$ populations for ages $< 8 \Gyr$ and low-$\alpha$ populations for ages $> 8 \Gyr$.

We make this cut in $\alpha$-element populations for two reasons.
First, the relative fractions of high- and low-$\alpha$ stars are almost always far from equal at any given age.
Given that the vast majority of high-$\alpha$ stars have ages $\gtrsim 8 \Gyr$ while most low-$\alpha$ stars exhibit ages spanning $0 - 10 \Gyr$, this cut provides a broadly sensible distinction.
That said, a significant number of stars seemingly depart from the simple heuristic that high-$\alpha$ stars are old and low-$\alpha$ stars are young.
Much debate has surrounded the nature of reportedly young (age $< 6 \Gyr$) high-$\alpha$ stars, with a general consensus that these stars are not genuinely young, but are instead the product of binary evolution, having gained mass through mergers or accretion \citep[for example][]{Miglio21, Jofre23, Grisoni24}.
On the other end, the nature of old ($> 10 \Gyr$) low-$\alpha$ stars remains unclear, potentially because the formation of these populations truly overlapped \citep{Silva18}, or because significant older high-alpha stars are subject to significant age uncertainties \citep{Haywood13}.
See \citet{Buder19} for further discussion.
Nevertheless, the mean $\sigma$ for stars with ages $\gtrsim 8 \Gyr$ is largely unaffected by this cut, with the inclusion or exclusion of old low-$\alpha$ stars changing the mean $\sigma$ by $\lesssim 5\%$.

In Figure~\ref{fig:mw obs}, all three components ($\sigma_{R}$, $\sigma_{\phi}$, $\sigma_{Z}$) exhibit a broadly similar age dependence: a steady rise for stars younger than $\approx$ 8 Gyr followed by a shallower slope at older ages.
This flattening is most apparent in $\sigma_{R}$ and $\sigma_{\phi}$, whereas $\sigma_{Z}$ continues to increase with age more noticeably, albeit more gradually than it does at younger ages. 
Across all ages, the radial component dominates, with the mean $\sigma_{R}$ exceeding the means of both $\sigma_{\phi}$ and $\sigma_{Z}$ by at least 50\%.
For stars younger than $\approx$ 8 Gyr, $\sigma_{\phi} > \sigma_{Z}$, while the two components become roughly comparable for older populations. These trends -- $\sigma_{Z}$'s continued rise, $\sigma_{R}$'s dominance, and $\sigma_{\phi}$ and $\sigma_{Z}$'s convergence at old ages -- are in remarkable agreement with the results of \citet{McCluskey} for FIRE-2 (see their Figure 4).
We further explore the physical drivers of the relative strength and evolution of each component of $\sigma$ in Section~\ref{subsubsec:fire mw}.

Despite the general agreement among the different observational datasets that each component of $\sigma$ increases with age, there is considerable scatter in both the normalization and slope of the relation.
Even after discounting potentially biased measurements from high- or low-$\alpha$ samples, the total 3D velocity dispersion at fixed age can vary by $\approx 25\%$, with $\sigma_{Z}$ alone varying by up to 50\%. Nevertheless, none of the observational trends show evidence of a factor of two increase in $\sigma$ over the age range.

\begin{figure*}
\centering
\hspace{-8 mm}
\includegraphics[width = 0.77 \textwidth]{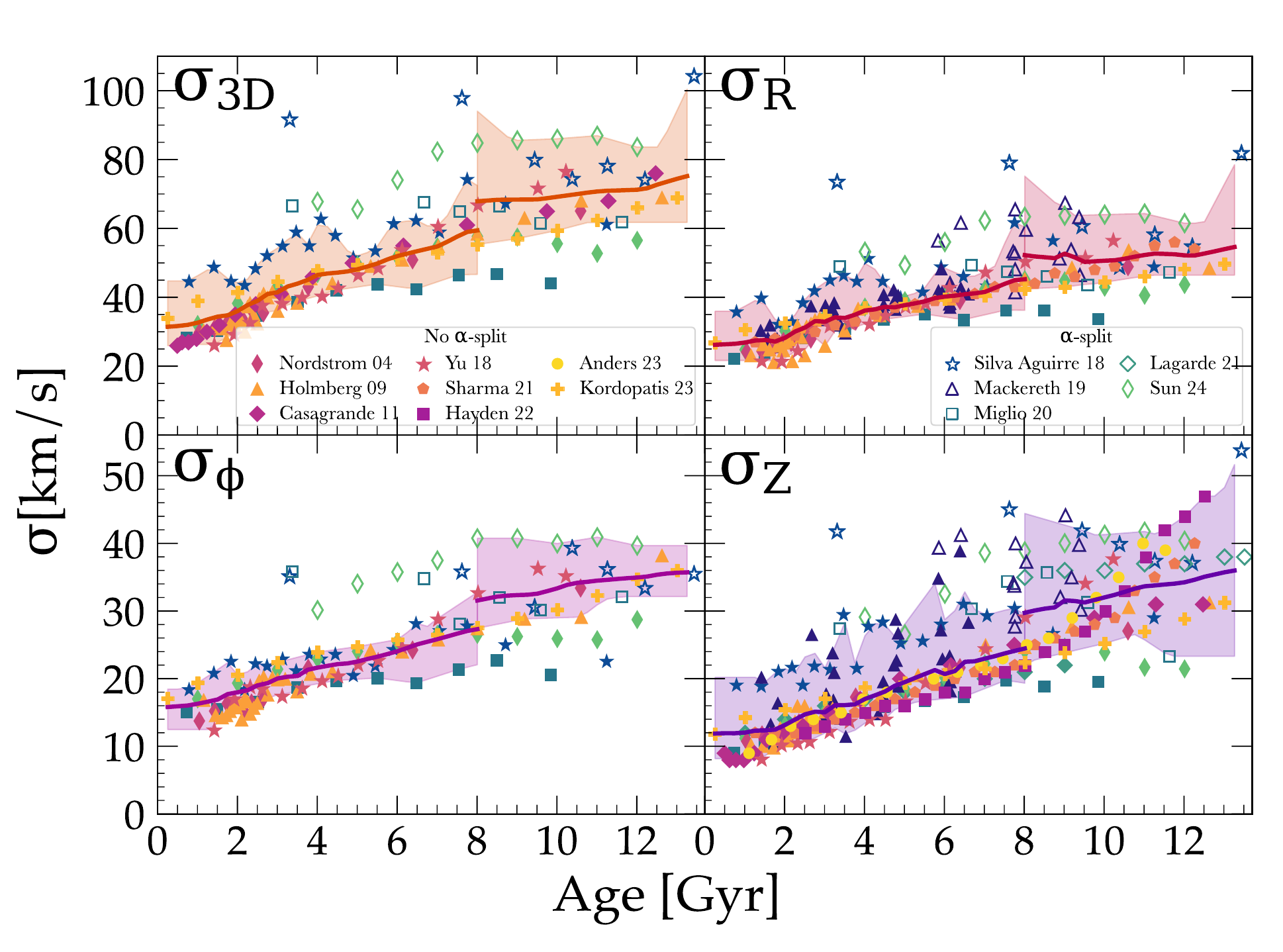}
\vspace{-2 mm}
\caption{
\textbf{Compilation of stellar velocity dispersion versus age observed for the Milky Way.}
``Warm" colored points show works that included all stars, while ``cool" colored points show works that separated stars into high-$\alpha$ (unfilled markers) and low-$\alpha$ (filled markers).
Solid lines show the mean, and shaded region shows the full range, across the observations, see \S\ref{subsec:mw observations} for details and caveats, including the origin of break in the mean relation.
All works agree that each component of $\sigma$ increases with age, but these works show substantial scatter for both the normalization and shape of the observed relation.
$\sigma_{R}$ unambiguously dominates: the mean $\sigma_{R}$ is at least 50\% greater than the mean $\sigma_{\phi}$ and $\sigma_{Z}$ at all ages. For ages $\lesssim$8 Gyr, $\sigma_{\phi} > \sigma_{Z}$, while $\sigma_{\phi} \approx \sigma_{Z}$ for ages $>$ 8 Gyr.
}
\label{fig:mw obs}
\end{figure*}

\newcolumntype{L}[1]{>{\centering\arraybackslash}p{#1}<{\centering\arraybackslash}}

\begin{table*}
\setlength{\tabcolsep}{0.2 cm}
\setlength\extrarowheight{2.3pt}
\caption{
Compilation of Observational Analyses of the MW in our Comparison
}
\centering
\begin{tabular}{|L{3.4cm}|L{2.6cm}|L{1.5cm}|L{2cm}|L{1.85cm}|L{1.4cm}|L{1.4cm}|}
\hline
Reference & Survey(s) Used & Stellar Type & Age Method & Spatial Selection [kpc] & Velocity Components & $\alpha$-element Split? \\
\hline
\rowcolor[gray]{0.9}[0.2 cm] \cite{Nordstrom2004} & GCS & F+G dwarfs & Isochrone & $d < 0.1$ & U, V, W  & - \\
\cite{Holmberg09} & GCS & F+G dwarfs & Isochrone & $d < 0.1$ & U, V, W & - \\
\rowcolor[gray]{0.9}[0.2 cm] \cite{Casagrande11} & GCS & F+G dwarfs & Isochrone & $d < 0.1$ & W, 3D & - \\
\cite{Silva18} & APOGEE DR10, Kepler, Gaia DR1 & RC/RGB & Asteroseismic & $d < 2$ ~ $|Z| < 1$ & U, V, W & yes \\
\rowcolor[gray]{0.9}[0.2 cm] \cite{Yu_18} & LAMOST DR3, Gaia DR1 & SGB/RGB & Isochrone & $d < 1$ ~ $|Z| < 0.27$ & $R$, $\phi$, $Z$ \newline & - \\
\cite{Mackereth19} & APOGEE DR14, Gaia DR2 & RC/RGB & Spectroscopic & & $R$, $Z$ & yes \\
\rowcolor[gray]{0.9}[0.2 cm] \cite{Lagarde21} & APOGEE DR14, Kepler, Gaia DR2 & RC/RGB & Asteroseismic & $R = 7.5 - 8.5$ $|Z| < 1.2$ & $R$, $\phi$, $Z$ & yes \\
\cite{Miglio21} & APOGEE~DR14, Kepler, Gaia~DR2 & RC/RGB & Asteroseismic & $d < 1-2$ & $Z$ & yes \\
\rowcolor[gray]{0.9}[0.2 cm] \cite{Sharma21} & GALAH~iDR3, Gaia~DR2 & MSTO  & Spectroscopic & $R = 6 - 10$ $|Z| < 0.6$ & $R$, $Z$ & - \\
\cite{Hayden22} & GALAH~DR3, Gaia~DR2 & All & Chemical & $R = 7.1 - 9.1$ $|Z| < 0.5$ & $Z$ & - \\
\rowcolor[gray]{0.9}[0.2cm] \cite{Anders23} & APOGEE DR17 & RGB & Spectroscopic & $R = 7 - 9$ & $Z$ & - \\
\cite{Kordopatis23} & Gaia DR3 & MSTO & Isochrone & $d < 1$ & $R$, $\phi$, $Z$ & - \\
\rowcolor[gray]{0.9}[0.2 cm] \cite{Sun24} & LAMOST DR4, Gaia EDR3 & RC & Asteroseismic & $R = 7 - 9$ $|Z| < 1$ & $R$, $\phi$, $Z$ & yes \\
\hline
\end{tabular}
\tablecomments{
All distances in kpc.
For the reported spatial selection, $d$ is the total 3D distance from the Sun, while $R$ and $Z$ refer to Galactocentric radial and vertical coordinates.
For the reported components of $\sigma$, 
$U$, $V$, and $W$ are Galactic space-velocities, where $U$ points towards the Galactic center, $V$ the direction of rotation, and $W$ the north Galactic pole, while $R$, $\phi$, $Z$ are Galactocentric cylindrical coordinates.
Works that report all 3 velocity components necessarily provide the 3D total dispersion as well (either directly or indirectly), but we do not include 3D in the listed reported components, with the exception of works that report the total dispersion sans the three components.
}
\label{table:table_mw}
\end{table*}

\subsection{Comparing Observed Galaxies}
\label{subsec:comparing observations}

\begin{figure}
\centering
\includegraphics[width = \columnwidth]{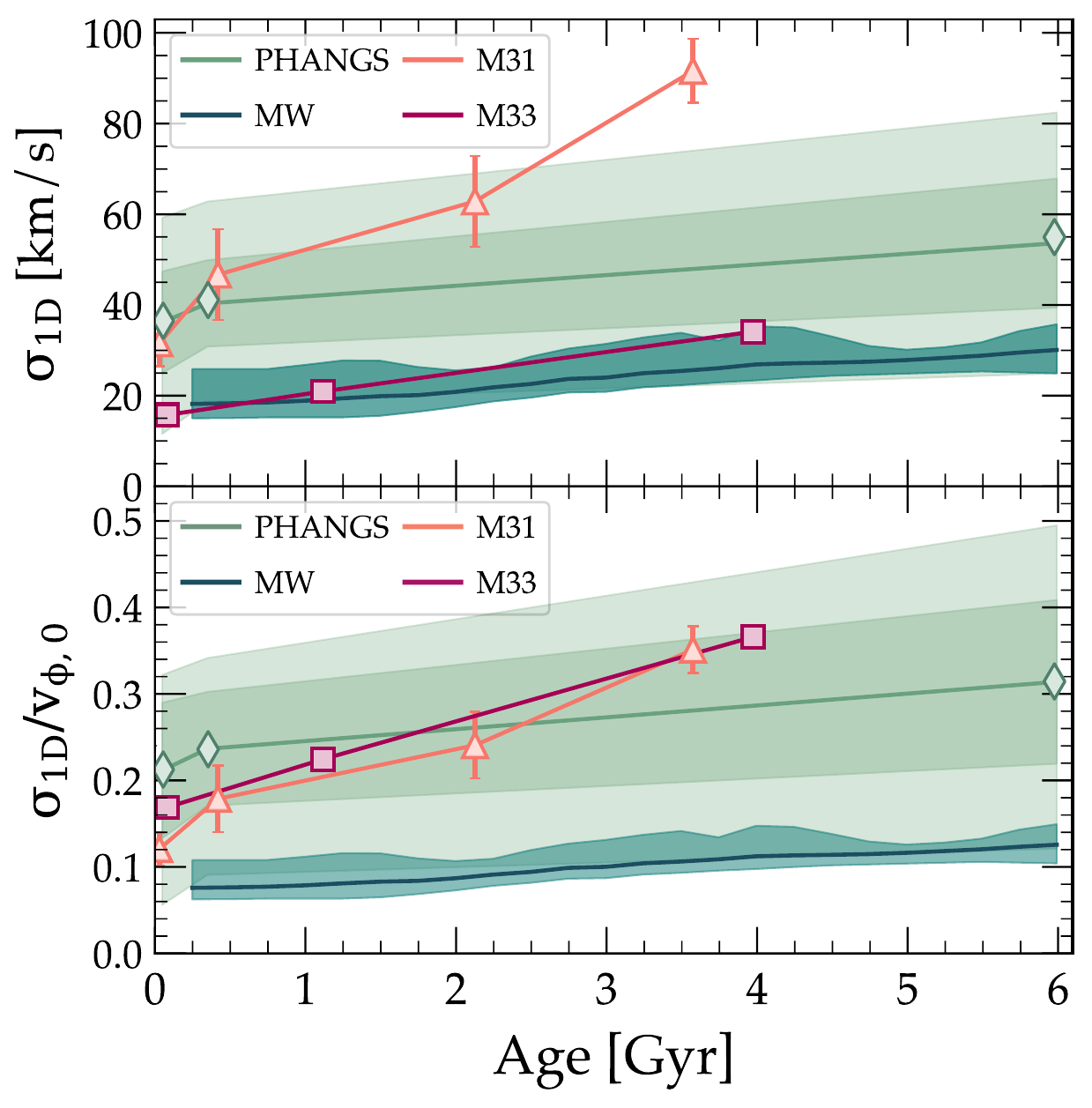}
\vspace{-2 mm}
\caption{
\textbf{Comparison of stellar velocity dispersion versus age in nearby galaxies.}
\textbf{Top}: The line-of-sight (1D) velocity dispersion of stars, $\sigma_{\rm 1D}$, versus age, as observed in the MW, M31 \citep{Dorman2015}, M33 \citep{Quirk22}, and 16 nearby galaxies from the PHANGS survey \citep{Pessa23}.
For the MW, the teal line and shaded region show the mean and full scatter across the observational works in Figure~\ref{fig:mw obs} for $\sigma_{\rm 1D} = \sigma_{\rm 3D} / \sqrt{3}$.
The light green line and shaded regions show the median, 68\% scatter, and full range across the PHANGS sample.
\textbf{Bottom}: Same, but scaled to each galaxy’s rotational velocity, $v_{\phi,0}$, for young stars today, to provide a dimensionless metric of dynamical coldness that is more comparable across these different-mass galaxies.
$\sigma_{\rm 1D}$ and $\sigma_{\rm 1D} / v_{\phi,0}$ increase with age for all galaxies and at all ages, \textit{though at differing rates with age}.
M31 and M33 have broadly similar $\sigma_{\rm 1D} / v_{\phi,0}$, which also agrees with (lies within the 68\% scatter) of PHANGS.
However, among this total sample of 19 galaxies, \textit{the MW is an outlier}, being dynamically colder in terms of $\sigma_{\rm 1D} / v_{\phi,0}$ than all but one galaxy from PHANGS.}
\label{fig:obs_all}
\end{figure}

Figure~\ref{fig:obs_all} shows $\sigma(\tau)$ for the MW, M31, M33, and PHANGS galaxies. For the MW, we show the mean and range of the 13 observations from Figure~\ref{fig:mw obs} (top left), and we convert the MW's total 3D $\sigma$ to a 1D value by assuming isotropy: $\sigma_{\rm 1D} = \sigma_{\rm 3D} / \sqrt{3}$. 
Some such conversion is necessary, given that the other observations show inherently 1D $\sigma_{\rm LOS}$. That said, stars in the MW are generally not isotropic but instead typically have $\sigma_{R} > \sigma_{\phi} > \sigma_{Z}$. However, we assume isotropy for this comparison to account for the wide range of inclinations of the other observed galaxies. Figure~\ref{fig:inc} shows that $\sigma_{\rm 3D} / \sqrt{3}$ typically matches $\sigma_{\rm LOS}$ at inclinations $45 - 50^{\circ}$, but over/underestimates $\sigma_{\rm LOS}$ at smaller/larger inclination. However, these estimates are typically within 25\% and 10\% of the proper value for young stars and old stars, respectively. In turn, inclination effects, or the faulty assumption of isotropy, cannot cause a factor of 2-3 difference between galaxies.  

Direct comparisons between galaxies are complicated by the different age binning employed by these works. 
First, data for external galaxies are limited to 3 or 4 age bins of unequal width, whereas most MW datasets use narrower and more uniform bins.
Second, the youngest age bins for external galaxies are $0 - 100 \Myr$ but are $250 - 500 \Myr$ in the MW. MW datasets rarely include the youngest ages easily probed in external galaxies, because MW observations are limited by the scarcity of young stars in the solar neighborhood and subject to different biases introduced by survey design and selection criteria. For example, GCS excludes blue stars entirely, whereas other surveys remove stars with poorly constrained ages, a common issue for younger RGB stars.
This gap between the youngest age bins in external galaxies and the MW can be partially addressed by including blue stars with imprecise ages, as in \cite{Aumer09}, who found $\sigma_{\rm 3D} / \sqrt{3} \approx 8 \kms$.
Alternatively, young star clusters provide well-dated populations, as demonstrated by \cite{Tarricq21}, who measured $\sigma_{\rm 3D} / \sqrt{3} \approx 5.2$ and $7.3 \kms$ for clusters aged 15 Myr and 100 Myr.
Including these other populations further suggests that the MW is particularly kinematically cold.

Figure~\ref{fig:obs_all} (top) shows $\sigma_{\rm 1D}$ versus stellar age.
The MW and M33 show significant overlap at all ages, and while some PHANGS galaxies reach similarly low values, most PHANGS galaxies have $\sigma$ that are twice as large.
Furthermore, although M31 falls within the 68\% scatter of the PHANGS sample at ages $\lesssim 1 \Gyr$, by 4 Gyr its $\sigma$ towers over that of the other galaxies, reaching values that are approximately $2 \times$ that of the typical PHANGS galaxy and $4 \times$ those of M33 and the MW. However, we caution against taking this comparison of $\sigma_{\rm 1D}$ at face value because it does not account for the order-of-magnitude range in these galaxies' stellar masses. 

Figure~\ref{fig:obs_all} (bottom) shows a more robust comparison, where we now scale each galaxy's measured $\sigma_{\rm 1D}(\tau)$ by its measured circular velocity today, $v_{\rm \phi,0}$, as obtained from the literature.
For M31 and M33, we adopt the maximum rotational velocity of the youngest stars from \citetalias{Dorman2015} and \citetalias{Quirk22}, respectively. 
For each PHANGS-MUSE galaxy, we use the peak velocity of the CO-based rotation curve, as determined in \citet{Lang20} using PHANGS-ALMA observations. 
Lastly, for the MW, we adopt the circular velocity at the solar position from the literature review of \citet{bhg-16}.
We list these $v_{\rm \phi,0}$ values, and their respective sources, in Table~\ref{table:table_obs}. 
Ideally, our adopted $v_{\rm \phi,0}$ values would rely on identical methods for each galaxy; that said, these different methods generally yield values within a few percent of each other, which would have little impact on our main conclusions.

Figure~\ref{fig:obs_all} (bottom) shows a clear contrast between the MW and nearby galaxies. External galaxies, although constrained to broad age bins with only partial overlap with MW data, exhibit $\sigma / v_{\phi,0}$ that are almost all higher than the MW. Compared to the MW, $\sigma / v_{\phi,0}$ is higher in M31 by $2.3 - 3.3 \times$ (depending on age), is higher in M33 by $2.4 - 2.8 \times$, and is higher than the average PHANGS galaxy by $2.7 - 3 \times$. Furthermore, 15 of 16 PHANGS galaxies exceed the MW’s value by at least a factor of 1.5, while the 68th percentile range for PHANGS is consistently double that of the MW. The MW's relation becomes even more striking when compared to the remarkable uniformity of M31, M33, and the average PHANGS galaxy, whose $\sigma / v_{\phi,0}$ vary by no more than 20\%, provided that we exclude the likely overestimated $\sigma$ at the youngest ages in PHANGS.

Although PHANGS galaxies exhibit a mean $\sigma / v_{\phi,0}$ that is more than double that of the MW, we emphasize that one galaxy (NGC 1433) has values nearly identical to the MW's.
This is an important benchmark, given the different methodologies of measuring $\sigma$ in the MW versus in nearby galaxies: NGC 1433 shows that at least one other disk galaxy has measurable stellar kinematics similar to that of the MW.
NGC 1433's low $\sigma$ may reflect (partially) the galaxy's low inclination ($28.6^\circ$), because, as Figure~\ref{fig:inc} shows, such inclinations lead to measured dispersions that are $10 - 25\%$ lower than isotropic predictions (which we assumed for the MW in Figure ~\ref{fig:obs_all}.
However, low inclination alone cannot explain NGC 1433's low dispersions, because other PHANGS galaxies exhibit similar, if not smaller, inclinations.
NGC 1433's properties provide further explanations for its cold kinematics. Its stellar mass, $7.9 \times 10^{10} \Msun$---the second highest in the PHANGS sample---aligns with theoretical predictions from \cite{McCluskey} that link higher galaxy masses with more vertically settled disks. Moreover, the galaxy’s subdued star formation rate of $1.1 \Msun {\rm yr}^{-1}$, comparable to the MW’s $\approx 2 \Msun {\rm yr}^{-1}$ \citep{Elia22}, places it 0.36 dex below the star-forming main sequence, the largest offset in the PHANGS sample. With lower SFRs comes lower stellar feedback and thus a less turbulent ISM, allowing stars to form with colder kinematics than those in more turbulent galaxies. These shared traits of NGC 1433 and the MW -- moderately high stellar masses and low SFR -- provide useful information in understanding how comparatively cold stellar kinematics relates to a galaxy's recent and cumulative star formation history.
We further discuss potential explanations for why the MW is kinematically colder than the majority of these observed galaxies in Section~\ref{subsubsec:mw outlier}.

\subsection{Comparing FIRE-2 with Observations}
\label{subsec:fire comparison}

We now compare the $\sigma(\tau)$ of these observed galaxies to those of FIRE-2 simulations.
We separately compare each observed galaxy (or galaxy sample in the case of PHANGS), so that we match the aperture size, radial selection, inclination angle, and age-binning of each work.
For all of these comparisons, we scale each (observed and simulated) galaxy by its $v_{\phi,0}$, which for FIRE-2 we define as the azimuthal velocity of stars with age $< 100 \Myr$ measured in apertures of $r_{\rm ap} = 250 \pc$ across the annulus at $R = 8 \kpc$, although these values exhibit little dependence on the exact age and aperture size.
We adopt age uncertainties of 20\% when measuring $\sigma(\tau)$ in FIRE-2, reflecting the lower end of typical observational age uncertainties.

\subsubsection{FIRE-2 versus M31, M33, and PHANGS}

We first compare FIRE-2 simulations to M31, M33, and PHANGS.

For our M31 comparison, we adopt an inclination angle of $77^\circ$ and an aperture radius of 760 pc -- the mean $r_{\rm ap}$ used by \citetalias{Dorman2015}.
The age bins in \citetalias{Dorman2015} have fairly broad and overlapping age distributions, but each binned distribution showed a distinct peak.
To compromise between ``broad and overlapping" and ``clearly defined" age bins, we use bins that span $0 - 0.1 \Gyr$, $0.1 - 1 \Gyr$, $0.5 - 3.17 \Gyr$, and $1 - 10 \Gyr$ (where $3.17 \approx \sqrt{10}$).
We measure $\sigma$ across annuli centered at $R = 5$, 8, 11, and 14 kpc broadly to match the radial range probed by \citetalias{Dorman2015} while accounting for the fact that our stellar disks do not extend as far as M31's and thus suffer from poor sampling at $R \gtrsim 16 \kpc$.

For our M33 comparison, we use an inclination of $56^\circ$, $r_{\rm ap} = 420 \pc$ (the median aperture in \citetalias{Quirk22}), and age bins of $0.02 - 0.18 \Gyr$, $0.56 – 2.2 \Gyr$, and $1.4 – 9 \Gyr$, corresponding to the $16 - 84^{\rm th}$ percentile ranges given for the age distributions in \citetalias{Quirk22}.
We measure $\sigma$ across annuli centered at $R = 1.5$, 4, 6.5, and 9 kpc, because these four equally spaced values bracket the radial range probed in \citetalias{Quirk22}.

For our PHANGS comparison, we use $r_{\rm ap} = 100 \pc$, age bins of 0-0.1, 0.1-0.6, and $3 - 10 \Gyr$ (the same as used by \citetalias{Pessa23}), and $R = 3$, 5, and 7 kpc. 
As before, we find the median $\sigma$ across the 3 annuli, which we then average across.
We repeat this for each of the 16 inclinations of the PHANGS sample ($8.9 - 57.3^\circ$).
We then average over these 16 measurements for each simulated galaxy, before ultimately computing the mean across our 11 simulations.

\begin{figure*}
\centering
\includegraphics[width = \textwidth]{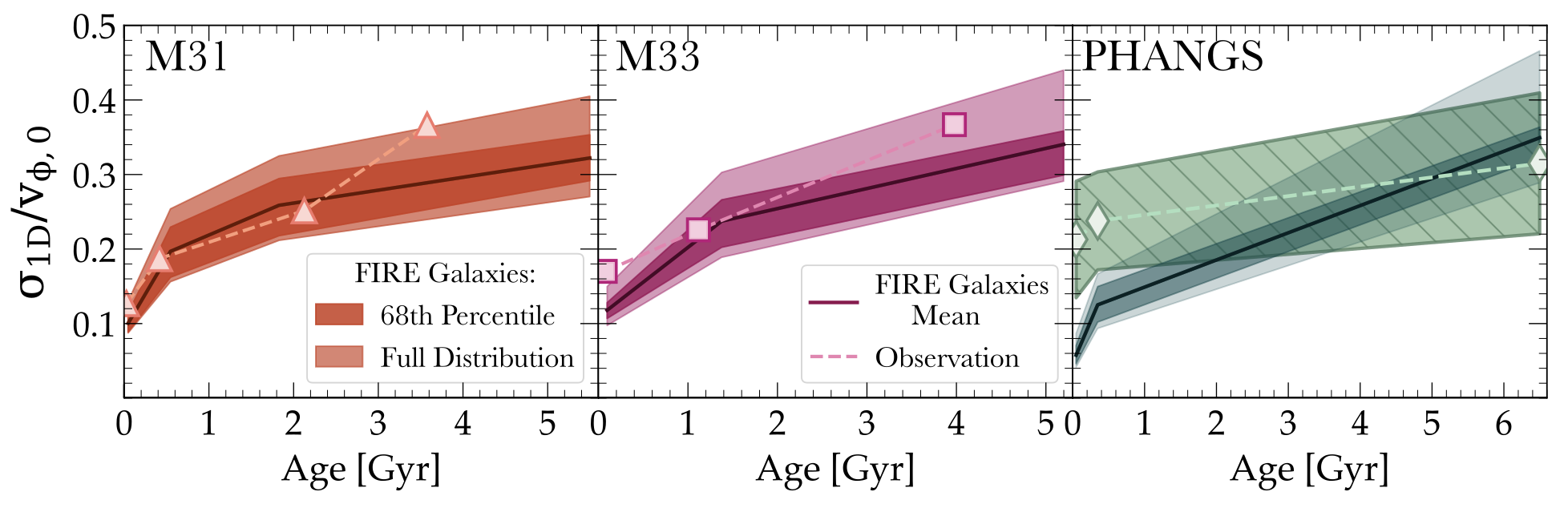}
\caption{
\textbf{Comparison of stellar $\sigma_{\rm 1D} / v_{\phi,0}$ versus age in the FIRE-2 simulations against observations of M31, M33, and 16 galaxies from the PHANGS-MUSE survey.} 
We emphasize that while $\sigma_{\rm 1D}$ depends on stellar age, $v_{\phi,0}$ is constant:
%$\sigma_{\rm 1D}$ increases with stellar age, but we divide by $v_{\phi,0}$ from only young stars today.
In measuring FIRE-2, we match the inclination angle, radial selection, spatial aperture size, and age binning of each corresponding observational sample, as in \S\ref{subsec:fire comparison}.
The solid lines and shaded regions show the median, 68 percent scatter, and full range scatter across 11 FIRE-2 galaxies.
The dashed lines with markers are from Figure~\ref{fig:obs_all} (bottom), showing observations of M31 \citep{Dorman2015}, M33 \cite{Quirk22}, and 16 galaxies from PHANGS-MUSE \citep{Pessa23}.
\textit{The trends for these 11 FIRE-2 galaxies are broadly consistent with those measured across this sample of 18 nearby galaxies}, with the key exception being young stars in PHANGS, which have \textit{higher} $\sigma$ than FIRE-2.
}
\label{fig:fire comp}
\end{figure*}

Figure~\ref{fig:fire comp} (left) compares observations of M31 from \citetalias{Dorman2015} with the M31-like apertures applied to FIRE-2.
FIRE-2 shows remarkable agreement with M31, especially at younger ages.
M31's last major merger event, $\approx 2.5 \Gyr$ ago \cite[for example][]{Hammer18, D'Souza18}, could not directly dynamically heat M31's youngest stars.
While M31's oldest age sample lies within the full distribution of FIRE-2, and presumably because of heating from the merger event, M31 falls at the high end of the FIRE-2 distribution.

Similarly, Figure~\ref{fig:fire comp} (middle) compares $\sigma(\tau)$ measured in FIRE-2 using ``M33-like apertures'' with the actual relation observed for M33 by \citetalias{Quirk22}.
Overall, FIRE-2 agrees well with M33, although the youngest stars in M33 exhibit slightly higher $\sigma$ than FIRE-2.
Considering the marked similarity between M31 and M33's $\sigma_{\rm LOS} / v_{\phi,0}$ in Figure~\ref{fig:obs_all}, it is perhaps not too surprising that FIRE-2 -- having just been seen to agree with M31 -- also agrees fairly well with M33. However, Figure~\ref{fig:obs_all} simply reports the values reported in the literature and does not account for the different radial ranges, aperture sizes, and inclinations of these observations.

Lastly, Figure~\ref{fig:fire comp} (right) compares the average (and sample scatter) of $\sigma(\tau)$ observed in the PHANGS-MUSE sample of 16 nearby galaxies with FIRE-2.
At first glance, the agreement is weaker than with M31 and M33, with the typical $\sigma_{\rm LOS} / v_{\phi,0}$ of the youngest stars being $4 \times$ higher in PHANGS than in FIRE-2.
However, this discrepancy at young ages likely results from the fact that the observed $\sigma_{\rm LOS}$ in PHANGS is overestimated, for two reasons.

First, the true velocity dispersion, $\sigma_{\rm true}$, of young populations is typically lower than the spectral resolution of the MUSE instrument, $\sigma_{\rm inst} \approx 62 \kms$ \citep{PHANGS-MUSE}.
If $\sigma_{\rm true} < \sigma_{\rm inst}$, the observed line profile is dominated by the spectrograph's line-spread function, effectively masking the intrinsic broadening of the narrower $\sigma_{\rm LOS}$. The tendency of Penalized Pixel Fitting (pPXF) to systematically overestimate $\sigma$ when $\sigma_{\rm true}$ is close to or near instrumental dispersions is well studied \citep[for example][]{Cappellari11, FalconBarroso17, vandeSande17, Varidel20}: various works found that this bias increases as $\sigma_{\rm true}$ decreases but largely disappears as $\sigma_{\rm true}$ exceeds the instrumental resolution.

Second, PHANGS measurements are not for individual stars, but for aggregate stellar populations within $\approx 100 \pc$ apertures.
For each aperture, the reported age is the luminosity-weighted average age of its constituent stars; moreover, because young stars dominate the light, these ages are biased by younger, brighter populations.
Thus, even apertures deemed very young could include a significant number of older stars.
In turn, the PHANGS measurement will overestimate $\sigma$ of ``young populations'', because even though older stars contribute negligibly to the light, they can contribute substantially to the line broadening \citep{Wang24}.
Together, these results indicate that $\sigma$ is typically overestimated by $\gtrsim 10 \kms$ relative to its true value in the youngest age bin, with the intermediate age bin likely showing weaker, but non-zero, systematic biases, while the oldest age bin is unaffected.
Therefore, one should take the observed values at the youngest ages shown in Figure~\ref{fig:fire comp} (right) with a grain of salt, and to the extent that they are lower in reality, they would agree better with FIRE-2.
At the oldest ages, where these instrumental effects are less important, FIRE-2 and PHANGS agree well.

Overall, $\sigma(\tau)$ in the FIRE-2 simulations exhibits remarkable similarity to those observed in M31 and M33, and is largely consistent with those in PHANGS galaxies, considering measurement effects.
This concordance goes against the standard lore that cosmological simulations have hotter kinematics than observed galaxies \citep[for example][]{House11}.
However, as we will show, this newfound harmony between (the stellar disk dynamics in) FIRE-2 and observations is not replicated for the MW.

\subsubsection{FIRE-2 versus the Milky Way}
\label{subsubsec:fire mw}

Finally, we compare the FIRE-2 simulations with the MW.
As we mentioned in Section~\ref{subsec:mw observations} and discussed in detail in Appendix~\ref{a:mw}, observational works measured the MW's $\sigma(\tau)$ using different stellar samples.
While some works measure stars in the immediate solar neighborhood ($d \lesssim 60 \pc$), others push towards the inner and outer disk ($d \lesssim 3 \kpc$).
We choose a middle ground for our comparison, measuring stars within apertures of $r_{\rm ap} = 250 \pc$ spanning an annulus centered at $R = 8 \kpc$.
%That said, our results for all but the youngest stars are insensitive to the aperture chosen, in agreement with the results of Section~\ref{subsec:ap_size}.
We adopt the same ``standardized'' age bins that we previously used to determine the mean (and range) among the assorted MW observations, that is, 250 Myr-wide age bins spanning $0.25 - 13.25 \Gyr$.
Furthermore, we also include a younger age bin ($0 - 0.25 \Gyr$) to highlight the rapid dynamical evolution of the youngest stars in FIRE-2 simulations.
That said, we caution against comparing results from this youngest FIRE-2 age bin with MW observations, given the lack of comparable MW data in this age range.
We note that unlike in previous sections, we do not compute the median $\sigma$ across the annulus of each galaxy.
Instead, we consider each aperture as a distinct measurement, which allows us to show the full range of ``solar neighborhoods'' in FIRE-2, not just the range of medians.

\begin{figure*}
\centering
\hspace{-8 mm}
\includegraphics[width = 0.77 \textwidth]{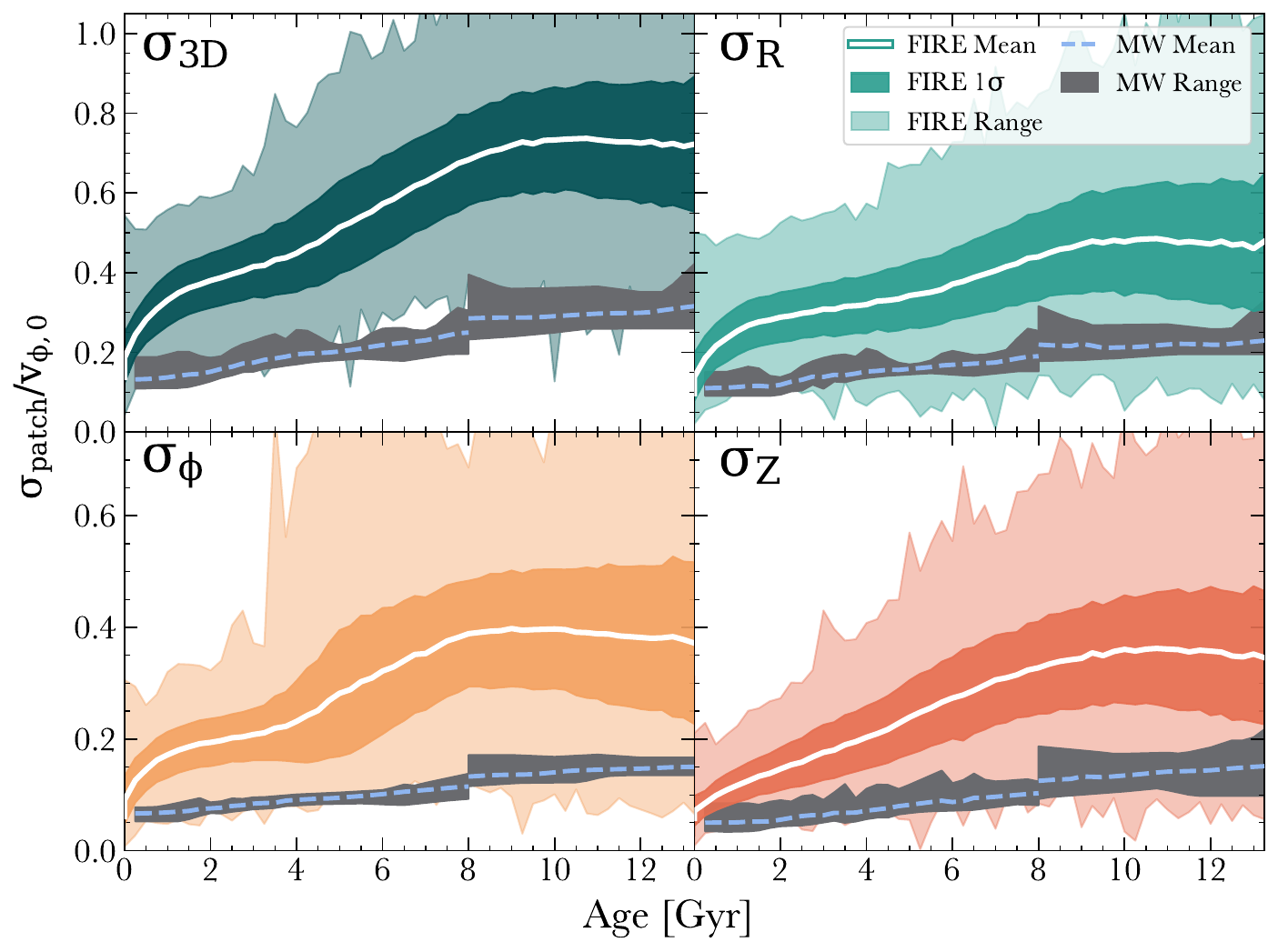}
\vspace{-2 mm}
\caption{
\textbf{Comparing stellar $\sigma(\tau) / v_{\phi,0}$ versus age, $\tau$, in the FIRE-2 simulations against observations of the MW} for $\sigma_{\rm 3D}$, $\sigma_{R}$, $\sigma_{\phi}$, and $\sigma_{Z}$. 
The solid white lines show the mean $\sigma(\tau) / v_{\phi,0}$, assuming 20\% age uncertainties, across all ''solar neighborhood" apertures in 11 simulations, while the dark and light shaded regions show the 68\% and full scatters across these apertures.
For the MW, we show the mean and full scatter of $\sigma(\tau) / v_{\phi,0}$ from the observational works in Figure~\ref{fig:mw obs}.
In general, FIRE-2 has significantly hotter stellar kinematics than the MW, with the exception of the youngest stars.
This offset is most pronounced for $\sigma_{\rm 3D}$ and $\sigma_{\phi}$, while $\sigma_{R}$ shows better agreement.
That said, these FIRE-2 galaxies do have some (rare) apertures that are as kinematically cold as the MW.
} 
\label{fig:mw fire}
\end{figure*}

In Figure~\ref{fig:mw fire}, we compare $\sigma(\tau) / v_{\phi,0}$ from the FIRE-2 simulations to the MW. Following the format of Figure~\ref{fig:mw obs}, we display the 3D, radial, azimuthal, and vertical velocity dispersion components separately. The FIRE-2 average appears in white, with dark and light shaded regions marking the 68\% and full simulation ranges, respectively. We overlay MW observations from Figure~\ref{fig:mw obs}, using dashed lines for the mean and gray shading to indicate the total observed scatter.

In general, all components show similar trends: the typical $\sigma(\tau) \ v_{\phi,0}$ is $\approx 3$ times larger in FIRE-2 than in the MW across all ages, except for the youngest stars, which suggests that post-formation dynamical heating is a driver the discrepancy between FIRE-2 and the MW.
However, comparing with young stars in the MW is difficult, because observational works have not yet measured ages and $\sigma$ for a large population of young stars \citep[though see][]{Zari23}.
Interestingly, some individual apertures at a given age agree with the MW.
However, no \textit{individual} aperture matches the MW across \textit{all} ages.
Having a low $\sigma$ at some ages does not necessarily indicate a low $\sigma$ at all ages for a particular region.

$\sigma_{R}$ agrees the best with the MW, followed by $\sigma_{Z}$ and then $\sigma_{\phi}$.
%, with $\sigma_{\rm 3D}$ showing the greatest discrepancy, especially for intermediate-aged stars.
This may reflect that the MW's thin disk began to form earlier than almost all galaxies in FIRE-2.
This means that at intermediate ages ($4 - 8 \Gyr$), MW stars likely formed and subsequently evolved in a thin disk, whereas in FIRE-2 these stars formed in a thicker and more turbulent disk.

\begin{figure*}
\centering
\hspace{-8 mm}
\includegraphics[width = 0.97\textwidth]{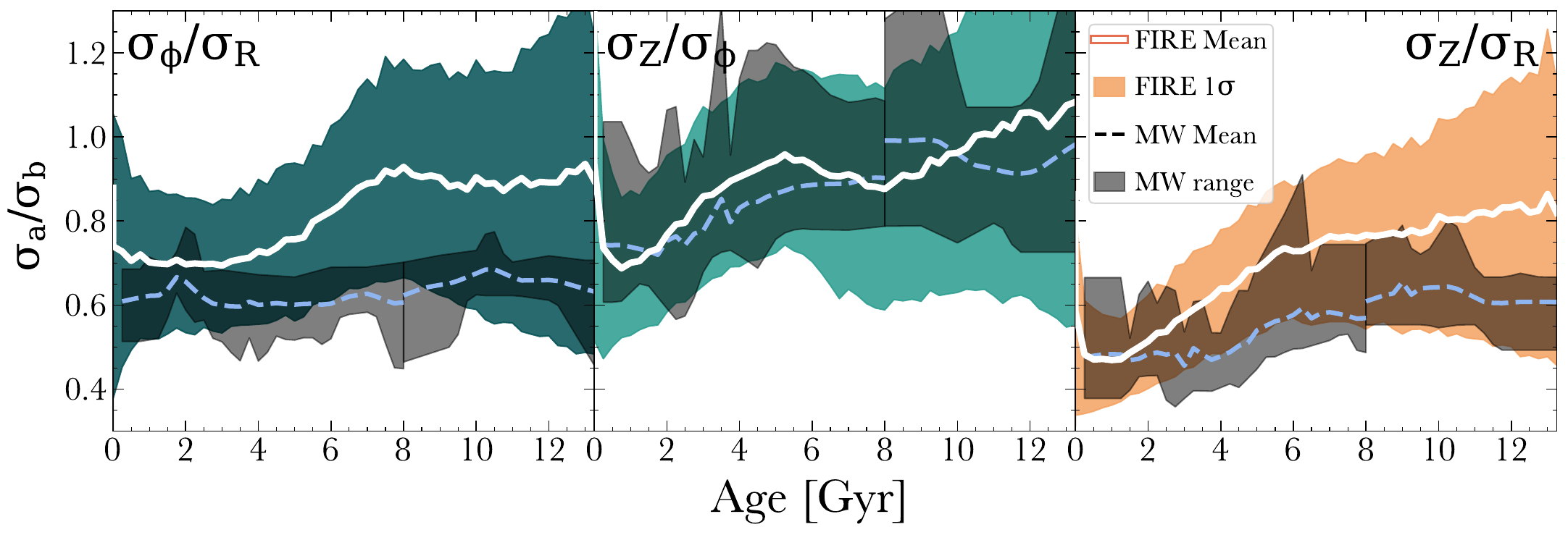}
\vspace{-2 mm}
\caption{
\textbf{Comparing the ratios of different components of the velocity dispersion versus age in FIRE-2 and the MW}.
Each white line shows the mean across all ``solar neighborhood'' apertures in the 11 simulations, and each shaded region shows the 68\% aperture-to-aperture scatter.
Dashed lines show the corresponding ratios for the MW, which we calculate using the mean $\sigma$ of the observational works in Figure~\ref{fig:mw obs}.
\textbf{Left}: In FIRE-2, the average $\sigma_{\phi} / \sigma_{R}$ exhibits different behavior for young ($\lesssim 4 \Gyr$), intermediate-aged ($\sim$ 4-8 Gyr), and old ($\gtrsim$ 8 Gyr) stars.
In contrast, $\sigma_{\phi} / \sigma_{R}$ in the MW remains flat with age at 0.6 until $\approx 9 \Gyr$ ago, before which it weakly increases.
The average value for MW lies below the mean from FIRE-2 mean, but the MW falls within the simulations' 68\% scatter at nearly all ages.
\textbf{Center}: $\sigma_{Z} / \sigma_{\phi}$ in FIRE-2 is consistent with the MW, growing from $\approx 0.75$ at 1 Gyr to $\approx 1$ at 13 Gyr in both.
\textbf{Right}: In both FIRE-2 and the MW, $\sigma_{Z} / \sigma_{R}$ generally increases with age, reflecting the growing importance of $\sigma_{Z}$ at older ages.
Although FIRE-2 and the MW agree at young ages, FIRE-2 exhibits a stronger increase with age at $1 - 6 \Gyr$, such that the mean in FIRE-2 is typically 30\% larger than the MW.
Nevertheless, the MW falls within the 68\% scatter in FIRE-2.
%at most ages, with the MW's $4 - 8 \Gyr$ population only narrowly falling outside of FIRE-2's scatter.
}
\label{fig:mw ratios}
\end{figure*}

We now examine the ratios of the components of $\sigma$, which are often considered to be key diagnostics of the relative contributions of different dynamical heating mechanisms
(\citealp{Smith12, Sellwood13}; though see \citealp{Pinna18, WaloMartin21}).
Figure~\ref{fig:mw ratios} shows three ratios -- $\sigma_{\phi} / \sigma_{R}$, $\sigma_{Z} / \sigma_{\phi}$, and $\sigma_{Z} / \sigma_{R}$ -- versus stellar age, comparing FIRE-2 (mean and $1 \sigma$ range) to the observed mean and range in the MW.
As mentioned, many MW studies either provide all 3 components or only $\sigma_{Z}$.
One study, \cite{Mackereth19}, included $\sigma_{Z}$ and $\sigma_{R}$ but not $\sigma_{\phi}$.
Therefore, our comparisons for $\sigma_{\phi} / \sigma_{R}$ and $\sigma_{Z} / \sigma_{\phi}$ include values from the 6 works that include all 3 components for the MW, whereas $\sigma_{Z} / \sigma_{R}$ includes values from these seven in addition to data from \cite{Mackereth19}.
We compute the ratio for each work, interpolate as discussed in Figure~\ref{fig:mw obs}, and then find the mean and range across all the works.
We confirmed that our results are not affected by the use of this subset of observational works, because this reduced sample yields average ratios nearly identical to those from the sample-averaged means of Figure~\ref{fig:mw obs}.

Figure~\ref{fig:mw ratios} (left) shows $\sigma_{\phi} / \sigma_{R}$ versus stellar age.
In the MW, $\sigma_{\phi} / \sigma_{R}$ is nearly constant with age, with a mean between 0.6 and 0.7.
In contrast, the mean in FIRE-2 shows a more complicated age dependence: $\sigma_{\phi} / \sigma_{R}$ is roughly constant at 0.7 between $0.25 - 5 \Gyr$, steadily increases from 0.7 to 0.9 between 5 and 8 Gyr, and finally remains constant at 0.9 for ages above 8 Gyr.
The fact that FIRE-2 is flat at the youngest epoch while the MW is flat across all ages likely reflects the dominance of the thin disk throughout the MW's history.
However, as the 1$\sigma$ scatter shows, some FIRE-2 galaxies have age dependence as flat as the MW.

For the case of nearly circular orbits, $\sigma_{\phi}$ and $\sigma_{R}$ are theoretically coupled via the epicyclic approximation, with a flat rotation curve predicting $\sigma_{\phi} / \sigma_{R} = \sqrt{2} / 2 \approx 0.71$ \citep{bt-08}.
Of course, the epicyclic approximation might not hold, because stellar orbits can be eccentric and/or the underlying gravitational potential can be non-axisymmetric or in disequilibrium.
However, we note that both for the entirety of the ages for the MW and for the youngest epoch for FIRE-2, the values are consistent with what is expected from the epicyclic approximation, whereas for older ages in FIRE-2, the approximation does not hold, as expected for stars on non-circular orbits.

Figure~\ref{fig:mw ratios} (center) shows $\sigma_{Z} / \sigma_{\phi}$ versus age.
Here, the averages for FIRE-2 and the MW are quite consistent.
In the MW, $\sigma_{Z} / \sigma_{\phi}$ increases from 0.75 to 0.9 for low-$\alpha$ stars and remains between 0.9 and 1 for high-$\alpha$ stars.
This initial increase and subsequent flattening also occur in FIRE-2, with $\sigma_{Z} / \sigma_{\phi}$ increasing from 0.7 to 1 for Late-Disk stars ($0 - 4 \Gyr$) and flattening/decreasing for Early-Disk stars ($4 - 8 \Gyr$).
The fact that these transitions occurred at younger ages in FIRE-2 ($\approx 4 \Gyr$) compared to the MW ($8 \Gyr$) probably reflects the earlier formation of the MW's disk, as discussed above.
Furthermore, unlike in the MW, FIRE-2 shows a third trend for the oldest stars, with $\sigma_{Z} / \sigma_{\phi}$ steadily rising from $0.9$ to $1.1$ for Pre-Disk stars.
This behavior for the oldest stars likely reflects the shorter Pre-Disk period of the MW and, correspondingly, the lack of Pre-Disk (halo) stars in the local sample.
However, we cannot exclude the possibility that FIRE-2 galaxies experienced stronger vertical heating in the Pre-Disk Era than the MW.
We also note that previous work found that this ratio is particularly sensitive to selection effects \citep{McMillan11, Aumer16}.

Lastly, Figure~\ref{fig:mw ratios} (right) shows $\sigma_{Z} / \sigma_{R}$ versus age.
In FIRE-2, $\sigma_{Z} / \sigma_{R}$ shows a strong increase with age, reflecting the increasing significance of vertical heating (relative to radial heating) over time.
This agrees with the results of \cite{McCluskey}, who found that the amount of post-formation vertical heating increased linearly with stellar age in these simulations, while in-plane heating plateaued at intermediate ages, and that $\sigma_{Z}$ had the largest relative impact from post-formation heating.
In comparison, the MW's trend is relatively flat, with only a modest increase with age.
This may indicate that FIRE-2 experienced more intense or extended vertical heating than the MW.

The ratio $\sigma_{Z} / \sigma_{R}$ has been studied extensively in the literature.  
\cite{Hanninen02} showed that it depends on the number density of GMCs, with larger densities yielding larger  $\sigma_{\rm Z} / \sigma_{\rm R}$. 
At early times, the higher mass and number density of GMCs rendered them particularly efficient heating agents.
However, this effectiveness decreased over cosmic time, because the relative mass contribution of GMCs decreased.
In contrast, large-scale spiral structure, being associated with the self-gravity of the stellar disc, increased in strength as the disk grew.
The fact that the MW exhibits lower values (and a flatter age dependence) may indicate that it sustained spiral structure for a larger fraction of its lifetime.
However, as with the left panel, the 1$\sigma$ scatter shows that some FIRE-2 galaxies have age dependence as flat as the MW.

Across all panels, the youngest stars exhibit large values for these ratios, as the rapid rise of the white line towards the left edge of each panel shows.
However, these values decline rapidly within a few 100 Myr, driven by a rapid increase in the in-plane components, particularly $\sigma_{R}$.
Specifically, in the first Gyr, $\sigma_{\rm 3D}$ more than doubles, increasing from 30 km/s to 75 km/s, with $\sigma_{R}$ accounting for 60\% of this heating, $\sigma_{\phi}$ accounting for 30\%, and $\sigma_{Z}$ accounting for just under 10\%.

In summary, these ratios of the components of $\sigma$ in FIRE-2 agree with the MW, in that the 1$\sigma$ scatter in FIRE-2 almost always overlaps with the mean of the MW.
However, the average FIRE-2 galaxy is higher than the MW for $\sigma_{Z} / \sigma_{R}$ and $\sigma_{\phi} / \sigma_{R}$, primarily because of the stronger increase with age in FIRE-2.

\section{Summary and Discussion}
\label{sec:discussion}

\subsection{Summary of Results}

We investigated the stellar age velocity (dispersion) relation, $\sigma(\tau)$, which is a key tracer of both stellar birth conditions (from the star-forming ISM) and subsequent dynamical heating of stars in disk galaxies.
We used a suite of 11 MW–mass galaxies from the FIRE-2 cosmological zoom-in simulations to assess how various observational effects influence the measured $\sigma(\tau)$, finding:

\begin{itemize} 
\item Age uncertainties of up to 40\% alter $\sigma$ by typically $\lesssim 20\%$ at any given age.
However, age uncertainties can significantly affect the slope of the relation between $\sigma$ and $\tau$, and age uncertainties $>$10\% can erase sharp features of major mergers.

\item For young stars (age $\lesssim 100 \Myr$), $\sigma$ increases significantly with the measured aperture radius, $r_{\rm ap}$, out to $> 1 \kpc$, reflecting turbulence in the star-forming ISM.
However, for stars older than this, $\sigma$ is essentially independent (to within a few percent) of aperture radius, at least at $r_{\rm ap} \gtrsim 250 \pc$, where FIRE-2 simulations spatially resolve $\sigma$ at that age.

\item $\sigma$ increases with decreasing galactocentric radius at all ages, by a factor of $1.5 - 2$ from $R = 12$ to $2 \kpc$.
The youngest stars have the strongest radial dependence: stars are born dynamically hotter towards the galactic center.
However, slightly older stars ($\approx 400 \Myr$ old) show the weakest dependence, likely from radial redistribution, while progressively older stars show somewhat stronger radial dependence.

\item Galaxy inclination affects measured line-of-sight dispersion, $\sigma_{\rm LOS}$, by up to a factor of $2$, with larger $\sigma_{\rm LOS}$ for inclinations $> 50 \degree$ compared to face-on observations.
This is because generally $\sigma_R > \sigma_\phi > \sigma_Z$, so a face-on disk generally leads to the smallest $\sigma_{\rm LOS}$.
This effect is the largest for fairly young stars ($0.4 - 2 \Gyr$) but weaker for newly-formed stars (ages $\lesssim 100 \Myr$ are affected up to 40\%).
Stars $\gtrsim 6 \Gyr$ old, being the most isotropic, are affected by at most 25-30\%.
\end{itemize}

We compiled observational measurements of $\sigma(\tau)$ from the literature: 13 analyses of the MW and analyses of M31, M33, and 16 PHANGS galaxies.
We compared them primarily in terms of $\sigma(\tau) / v_{\phi,0}$, where $v_{\phi,0}$ is the rotation velocity today, because this dimensionless ratio provides the most robust way to compare these galaxies that span over an order of magnitude in stellar mass.
We found:

\begin{itemize}

\item All analyses of the MW show an increase in $\sigma$ with age for all velocity components ($v_R$, $v_\phi$, $v_Z$), by a factor of $2 - 3$ over the last $\approx 13 \Gyr$.
However, these works show substantial scatter ($\approx 50\%$), in both the normalization of $\sigma$ at fixed age, and shape of the relation of $\sigma$ with age.
Within the MW, $\sigma_{R}$ dominates, being at least 50\% greater than $\sigma_{\phi}$ and $\sigma_{Z}$ at all ages.

\item M31, M33, and the average PHANGS galaxy exhibit remarkably consistent values of $\sigma(\tau) / v_{\phi,0}$.
M31 and M33 are nearly identical and generally within the 1$\sigma$ range of the PHANGS sample.

\item Thus, among this sample of 19 nearby disk galaxies, the MW is a strong kinematic outlier.
The MW is dynamically colder than M31, M33, and 15 of the 16 PHANGS galaxies.
The MW is similar to one PHANGS galaxy, NGC 1433, so importantly, it is possible to measure at least one other nearby galaxy as dynamically cold as the MW>
\end{itemize}

Finally, we compared $\sigma(\tau) / v_{\phi,0}$ in the FIRE-2 simulations against these observations, accounting for the observational effects that we explored.
We found:

\begin{itemize}
\item FIRE-2 galaxies show good agreement with M31 and M33 across all measured ages.
FIRE-2 matches PHANGS galaxies at older ages, but at the youngest ages ($\lesssim 100 \Myr$), $\sigma / v_{\phi,0}$ is lower in FIRE-2 by a factor of 4.
However, this discrepancy is likely not physical, but instead reflects the (well-known) limitations of measuring small $\sigma$ with MUSE.

\item FIRE-2 galaxies show $\sigma(\tau) / v_{\phi,0}$ typically $2 - 3 \times$ higher than the MW at a given age.
The youngest stars show the best agreement, indicating that post-formation dynamical heating is more rapid in FIRE-2 than in the MW.

\item The ratios of components of $\sigma$ in FIRE-2 agree with the MW within the 1$\sigma$ scatter, though there are offsets in the average relations.
In both FIRE-2 and the MW, $\sigma_R > \sigma_\phi > \sigma_Z$ at essentially all ages.
\end{itemize}

\subsection{Discussion}

\subsubsection{The Milky Way as an outlier among observations}
\label{subsubsec:mw outlier}

One of our most important results is that the MW is a kinematic outlier, with an unusually low $\sigma / v_{\phi,0}$ by a factor of $2-3$ at all measured ages, as compared with observed nearby disk galaxies. A possible concern is systematics in comparing measurements of $\sigma$ based on individual stars in the Solar neighborhood of the MW to $\sigma$ measured across many apertures of nearby galaxies.
However, our exploration of measurement effects in Section~\ref{sec:P1}, using FIRE-2 simulations, indicates that these effects alone are unlikely to be strong enough to explain why the MW is a kinematic outlier.
We also emphasize that one galaxy (out of 16) from the PHANGS-MUSE sample, NGC 1433, has $\sigma(\tau) / v_{\phi,0}$ similar to the MW. Thus, measuring $\sigma$ as low as in the MW is \textit{possible} in nearby galaxies, despite our external view. That is to say, the existence of a similarly cold galaxy actually strengthens the classification of the MW as a kinematic outlier. 

Furthermore, recent observational works arrive at similar conclusions regarding the MW's typicality. 
\cite{Bershady24} estimated $\sigma(\tau)$ for 500 star-forming disk galaxies in the MANGA survey by parameterizing their observed asymmetric drift, $\sigma_{\rm AD}$, which is the lag between the tangential speeds of stars and gas. \cite{Bershady24} also derived analogous $\sigma_{\rm AD}$ relations for the MW, M31, and M33 from literature data, finding that while M31 and M33 are broadly consistent with the typical trends in MANGA, the MW falls far below the standard relation.
These findings also provide further insight on the origin of the MW's discrepancy, because \cite{Bershady24} parameterized their exponential heating model via both the radial $\sigma$ of young stars and a heating index or ``stratification rate''.
These results indicate that although young stars in the MW are atypically cold, with similarly-massive MANGA galaxies exhibiting radial $\sigma$ nearly 3 times higher -- the \textit{rate} of heating in the MW (over the past few Gyr) is well within the range of observed values. 

The unusual stellar kinematics of the MW likely relates to its unique history, mainly, its early forming disk and paucity of major mergers. As we discuss below (and tritely summarize now), these factors allowed the majority of the MW's stars to form  ``disky'' and subsequently stay (reasonably) ``disky'' throughout the Galaxy's lifetime.

The earliest epoch of the MW's history has been the subject of extensive recent study \citep[for reviews see][]{Helmi21, Deason24}.
Such studies primarily focus on the orbit and abundance distributions of the Galaxy’s oldest, and thus, most metal poor, stars.%; however, many metal-poor stars formed ex situ, and thus do not represent the kinematics of the MW-progenitor. That said,  
Notably, \cite{Belokurov22} identify a large sample of in-situ stars from APOGEE DR17 and Gaia EDR3 that span iron abundances down to [Fe/H] $= -1.5$, and show that the mean $v_\phi$ of these stars rapidly increased from [Fe/H] of $-1.2$ to $-1$. They argue that this increase traces the MW's transition from a protogalaxy with negligible net rotation to a coherent disk with $v_{\phi} \approx 150 \kms$ within $1 - 2 \Gyr$. Subsequently, \cite{Conroy22} used the H3 survey to extend this analysis to even lower metallicities, and by measuring the ages of these stars, argued that the disk formed $\approx 10 - 12 \Gyr$ ago. On the other hand, \cite{Xiang22} and \cite{Xiang25} studied 250,000 subgiant stars in LAMOST (with a remarkably low typical age uncertainty of $\approx 8\%$) and concluded that the MW's disk emerged $\gtrsim 13 \Gyr$ ago, formed a total stellar mass of 3.7$\times10^9$M$_{\odot}$ by 13 Gyr ago, and then further experienced a 2-3 Gyr period of intense star formation.   

Indeed, the exact timing of the formation of the MW's disk remains rather uncertain \citep[for example][]{Viswanathan24, Zhang24_vmp, Horta25}. For example, age uncertainties of 10\% can shift inferred disk formation times to 1-2 Gyr earlier than in reality \citep{Zhang24_bar}, and accreted stars can masquerade as in-situ disk populations \citep{Santistevan21, Mardini22}.
Furthermore, the stellar orbits that we measure today do not necessarily reflect those at birth, especially for the oldest stars: \citet{McCluskey} showed that stars need not form with disk-like kinematics to exhibit significant coherent rotation today, given possible subsequent torquing by mergers, satellite interactions, or bars \citep[see also][]{Chandra24}.
Nevertheless, most works agree that the MW's (long-lived) disk probably formed no later than $9 - 10 \Gyr$ ago.

This places the MW at the earliest end of the distribution of disk formation, at least within cosmological simulations.
Only $\approx 10-13\%$ of the MW-mass analogs in the TNG50 and Artemis simulations form disks in stars at such low metallicities \citep{Dillamore24, Semenov24_mw1}, and this is only 6\% among the E-MOSAICS simulations \citep{Kruijssen19}. We note that some high redshift simulations form coherent disks even earlier, at $z = 4 - 7$ \citep{Tamfal22, Kohandel24, vanDonkelaar24}; however, many of these are already massive at these early times, with $\Mstar \gtrsim 10^{10} \Msun$, and thus are not representative of the typical MW-mass progenitors. Furthermore, when comparing against the MW and nearby galaxies, it is critical to compare with simulations that have been run to the present day and thus include the effects of long-term dynamical heating of stars.

The MW's early disk formation likely reflects its rapid early mass assembly.\footnote{Theoretical works indicate that galactic disks tend to emerge once their host halos exceed $\approx 10^{11} \Msun$ and/or a stellar mass of $\approx 10^{9-10} \Msun$) \citep{el-badry18, Pillepich19, Hopkins23}, although the exact link between mass assembly and disk formation remains debated \citep{Dekel09, Stern21, Hafen22, Semenov24_mw2}.}
Works based on the MW’s reconstructed SFH \citep{Snaith14} and globular cluster properties \citep{Kruijssen19, Trujillo-Gomez21} consistently predict that our Galaxy assembled 50\% of its z=0 stellar mass by 10-10.5 ($\pm$ 1.5) Gyr ago. In contrast, typical MW-mass halos reached 50\% of their final stellar mass $2 - 3 \Gyr$ later than this, based on semi-empirical models \citep{Behroozi19}, abundance matching of observed galaxy populations \citep{Papovich15}, and reconstructing star-formation histories of low-redshift galaxies \citep{Pacifici16}.

Once a disk forms, it tends to self-regulate near Toomre $Q \approx 1$, enabling subsequent generations of stars to form with progressively colder kinematics, provided its gas fraction declines (as expected) \citep[for example][]{Gurvich20}. For example, \citet{McCluskey} showed that FIRE-2 galaxies with earlier disk formation times tend to \textit{form} stars with colder present-day kinematics than those with later-forming disks \citep[see also][]{Semenov24_mw1}.

Forming a disk is only half the battle; preserving it is equally critical. Once disks have formed, mergers act to dynamically heat existing stellar populations, potentially reorienting and even destroying pre-existing disks \citep[for example][]{Semenov24_he}. In the limit that disks are maintained, simulations consistently show that mergers induce noticeable increases in stellar velocity dispersion and imprint discontinuities in $\sigma(\tau)$ \citep{Martig14, Grand20, Khoperskov23}. The MW likely underwent a major merger $\approx 10 \Gyr$ ago \citep{Helmi18, Belokurov18}. Since then, its merging history has been atypically tranquil for a galaxy of its present-day mass \citep{Hammer07}: for example, \cite{Evans20} found that only 5\% of MW-mass galaxies in the large-volume EAGLE simulation experienced an early major merger with no subsequent massive mergers since z=1 (which reduces to 0.65\% if an LMC-like satellite is also required). \footnote{The MW’s atypically rapid early assembly and similarly atypical subsequent calm are likely two sides of the same coin: galaxies with rapid early mass assemblies that then go on to experience more typical (less quiescent) merger histories wind up more massive than the MW, whereas galaxies with similarly minor late-time mass growth cannot reach MW masses unless they also reached similarly high masses early on.}

The MW’s unique formation history likely relates to its group environment, because galaxies in proto-group (that is, cosmically dense) environments experience more rapid mass accumulation at early times \cite[e.g.,][]{Gallart15, Santistevan20}. Furthermore, the LG’s unique cosmographic landscape and quiet Hubble flow likely facilitated the MW’s early disk formation and quiescent merger history, as various works find earlier formation and last major merger times for simulations constrained to match the LG’s larger environment \citep[for example][]{Forero-Romero11, Carlesi20}. These constrained simulations also help us to understand the differences between the MW and M31, since even though M31-mass  and MW-mass galaxies formed similarly quickly at early times, M31 analogues exhibited more late-time mass growth and essentially shielded their lower-mass (MW-analogue) neighbors from later mergers \citep{Wempe25}.

These theoretical works are in agreement with our understanding of both the MW and M31's merger histories. M31 appears to have experienced a recent major merger -- possibly as significant as a 1:4 mass ratio -- approximately 2.5 Gyr ago \citep{Hammer18, D'Souza18, Bhattacharya23, Tsakonas2024}. That said, M33’s merger history is more ambiguous. Its warped stellar and HI disks, coupled with a burst of star formation in both M31 and M33 roughly 2–3 Gyr ago, have long been interpreted as signatures of a past fly-by interaction with M31 \citep{McConnachie09, Bernard12, Williams17, Corbelli24}. However, recent observations and orbital modeling strongly suggest that M33 is on its \textit{first} infall into the Local Group and has yet to undergo a close pericentric passage with M31 \citep{Patel17, vanderMarel19, Patel25}. 

Lastly, we note that the MW's cold kinematics, early formation, and quiescent merger history are only a few of its potentially atypical properties.
For example, many works argue that the MW is abnormally compact for its stellar mass, having a remarkably short disk scale radius \citep{RixBovy13, Licquia16, bhg-16}.
However, recent work argues that the MW’s size is actually typical, and that its apparent compactness reflects the faulty assumption that its disk follows a simple exponential profile \citep{Lian24}.
When \citep{Lian24} fitted the MW's surface brightness profile nonparametrically, the Galaxy's measured half-light radius doubles, and its disk size is consistent with the overall galaxy population. However, this increased disk size makes the MW even more of an outlier in terms of its (abnormally) steep radial gradient in metallicity \citep{Boardman20, Lian23}.

\subsubsection{Caveats about FIRE-2 simulations}

We next discuss some key limitations of the FIRE-2 simulation and other caveats to our results.

$\sigma(\tau)$ in FIRE-2 agrees well with M31 and M33, partially with PHANGS galaxies, and is systematically higher than the MW.
We emphasize that the FIRE-2 simulations are not designed to recreate any individual galaxy, including the dynamically cold MW.
FIRE-2 represents halos (or LG-like pairs of halos) selected based only on their dark-matter mass at $z = 0$, agnostic to the halos' additional properties and formation/merger history.
Thus, the FIRE-2 results should represent \textit{random and representative} histories of MW-mass galaxies.
As such, it may not be surprising that among 11 FIRE-2 MW-mass galaxies, none of them agrees well with the normalization of $\sigma(\tau)$ in the MW, while they age better with larger samples of observed galaxies.
As \citet{McCluskey} discussed, and consistent with our discussion in the previous subsection, the LG-like FIRE-2 simulations tend to form disks earlier (in particular, Romeo), so they tend to have lower $\sigma(\tau) / v_{\phi,0}$ today and thus among the sample are closest to the MW.

That said, the FIRE-2 simulations in this paper do not model all the physics that could be relevant.
These simulations do not model magnetohydrodynamics (MHD) nor self-consistent cosmic-ray injection, transport, and feedback. \cite{Hopkins20cr} show that including cosmic rays in these simulations can lower the stellar mass (growth) of the galaxy, especially at late times, although the strength of the effect depends sensitively on the details of the assumed diffusion/transport of cosmic rays.
Cosmic-ray feedback can reduce a galaxy's present-day stellar mass and SFR, which can lead to a lower $\sigma$ in gas \citep{Chan21}.
However, \cite{McCluskey} show that this also leads to a slight \textit{reduction} in the degree of rotational support of these galaxies, in terms of $v_\phi / \sigma$, because this shift occurs primarily along the existing stellar mass-kinematics relation \citep[see Figure~A2 in][]{McCluskey}.
Thus, cosmic-ray feedback, as implemented in FIRE-2, does not make these disks meaningfully dynamically colder.

Potentially more important is that these FIRE-2 simulations do not model supermassive black holes and feedback via active galactic nuclei (AGN), although some recent FIRE simulations have \citep{Wellons23, Mercedes-Feliz23}.
Although AGN feedback is most critical for more massive galaxies, growing work finds that AGN can affect the structural and dynamical properties of MW-mass galaxies \citep[for example][]{Irodotou22}.
The lack of AGN feedback may explain how these FIRE-2 galaxies display high average SFR: the average SFR in these FIRE-2 simulations today is $\approx 5-10 \Msun$ yr$^{-1}$, higher than most (although not all) observational inferences at these masses \citep{Gandhi22}. In particular, the central regions of FIRE-2 galaxies display intense star formation stemming from strong gas inflows.
In real galaxies, these gas inflows would also feed the central black hole, and the resulting AGN feedback would regulate further inflows and star formation in the central regions. Indeed, \cite{Marasco25} argue that the lack of AGN feedback in FIRE-2 causes them to exhibit more complex and broader HI line profiles than in most observed galaxies.

Given the intimate connection between bar formation and gas dynamics in the inner galaxy \citep[for example][]{Athanassoula13, Fraser-McKelvie20, Fragkoudi25}, the lack of AGN feedback may also affect bars in these simulations.
For example, high gas fractions and star formation in the inner galaxy can drive fluctuations of the gravitational potential that disrupt bars as they form \citep{Weinberg24}.
Bars do form in about half of these FIRE-2 simulations, but they are generally weaker, more ellipsoidal, and shorter-lived than in observed galaxies \citep{Ansar25}.
Bars can affect stars throughout the galaxy and can be a driver of post-formation dynamical heating, either directly \citep[for example][]{Grand16}, or through resonant overlap with spiral structure \citep{Minchev10, Marques25}. 
As such, stronger bars might increase $\sigma$, especially $\sigma_{\phi}$ \citep{Asano22, Khoperskov22, Dillamore23, Filion23}.

Lastly, much of our analysis focused on quantifying the impact of measurement effects -- age uncertainties, aperture radius, galactocentric radius, and inclination angle -- applied to the ``truth'' of the FIRE-2 simulations.
%, i.e., selecting star particles within a given aperture, taking the velocity component relative to a given inclination angle, and then calculating the resulting velocity dispersion.  
However, we did not use synthetic observations that modeled observable photometry and spectroscopy, including magnitude limits, dust, etc.
In future work we will explore in more detail how both synthetic measurements of resolved stars and of IFS-like synthetic measurements of stellar population affect these results. 

\subsubsection{Comparison to Other Simulations}

We now discuss our results from the FIRE-2 simulations in relation to other cosmological zoom-in simulations of MW-mass galaxies.
We identified six other studies reporting $\sigma(\tau)$, spanning 48 MW-mass galaxies: 16 from Auriga \citep{Grand16}, 19 from {\small RaDES} \citep{Few12, Ruiz-Lara16}, five from {\small NIHAO-UHD} \citep{Buck20}, one from {\small VINTERGATAN} \citep{Agertz21}, one (h277) from the g14 suite \citep{Bird-21}, and 6 from {\small HESTIA} \citep{Khoperskov23}. 
We also included results for 11 FIRE-2 galaxies.
\textit{A critical point that we emphasize: among these 59 cosmological zoom-in simulations of MW-mass galaxies across seven suites, only a single galaxy, h277, matches the MW's $\sigma(\tau)$.
In contrast, these simulated galaxies generally agree better with the larger sample of M31, M33, and PHANGS.
}

We first note important caveats to this statement.
As mentioned, we limit this comparison to \textit{cosmological zoom-in simulations of MW-mass galaxies}, meaning we do not include results for isolated (non-cosmological) simulations \citep[for example][]{Kumamoto17, vanDonkelaar22}, non-hydrodynamic simulations, including those using sticky-particle algorithms \citep[for example][]{Martig14}, larger-volume simulations \citep[for example][]{Pillepich19, Pillepich24}, or simulations targeting significantly different masses than the MW \citep[for example][]{Brook12}.
Second, the galaxies for which the authors measure $\sigma(\tau)$ in the above simulation suites sometimes represent only a subset of the entire suite.
{\small HESTIA} features 13 LG-like pairs \citep{Libeskind20}, Auriga encompasses 30 galaxies \citep{Grand17}, and the g14 suite includes about a dozen otherwise well-studied galaxies \citep[for example][]{Loebman12, Munshi13, Kassin14, Christensen16, Brooks17}, of which h277 has by far the smallest $\sigma(\tau)$.

\textit{Also striking is that nearly all of these different simulation suites (FIRE-2, Auriga, NIHAO-UHD, VINTERGATAN, HESTIA, RaDES) show remarkable quantitative consistency among them for their average $\sigma(\tau)$}, all of which increases (generally monotonically) with age.
Most of these 59 galaxies have $\sigma_{Z} \approx$ 15 – 20 km/s for the youngest stars and 60-70 km/s for stars aged $\approx 6 \Gyr$, with some galaxies scattering to higher values.
This \textit{quantitative} agreement may be surprising, given the diversity of models for the ISM, star formation, and feedback used across these simulation suites and their diversity of initial conditions, and thus formation histories. 
These $\sigma_{Z}$ values generally match M31, M33, and (at least for stars older than $\approx 1 \Gyr$) PHANGS.
In contrast, $\sigma_{Z}$ in the MW is only $\approx 12$ and 20 km/s at the same ages.
Therefore, tension with the MW is somewhat stronger at older ages, which may reflect the MW's quiescent merger history and early-forming disk.
However, these explanations do not obviously apply to younger stars, which in almost all simulations formed in non-merging, rotationally-dominated disks.

The key exception is h277, which, as \cite{Bird-21} show, exhibits a $\sigma(\tau)$ similar to that of the Solar neighborhood of the MW.
Furthermore, as \cite{Bird-21} noted, h277 is unusual among the larger g14 simulation suite of over a dozen galaxies.
\cite{Brooks09} introduced h277, and \cite{Zolotov09} notes its particularly quiescent merger history.
However, the original version of h277 did \textit{not} stand out kinematically: \cite{House11} showed that it had the highest $\sigma_{Z}$ ($\approx 28 \kms$) among young stars in their simulated sample.
The updated version of h277 analyzed in \cite{Bird-21} had twice better spatial resolution (173 pc), eight times better mass resolution ($5800 \Msun$), and updated ISM and star formation models from \cite{Christensen12}, which directly link star formation to $\rm H_{2}$.

\cite{Bird-21} argues that the low $\sigma_{Z} \approx 7 \kms$ today arose primarily from their updated star-formation prescription, which limits star formation to dense gas ($n > 100$ cm$^{-3}$).
The connection between star formation thresholds and $\sigma$ of stars at birth may account for unrealistic kinematics in earlier simulations \citep{House11}.
However, the relationship between high-density star formation thresholds and low $\sigma$ in newly formed stars is not clear across different simulations.
For example, the FIRE-2 simulations we analyze here adopt a minimum density for star formation of $n = 1000$ cm$^{-3}$, along with a requirement of gas being self-gravitating.
An alternative expectation is that higher density thresholds combined with a self-gravitating criterion lead to more clustered star formation, which increases the impact of stellar feedback and dynamical perturbations on short timescales \citep{Fielding18, Martizzi20, Orr22}.
An even broader explanation is that while stars form from the densest self-gravitating gas, this gas is still not as kinematically cold as the gas in the MW (or h277).

The cold kinematics of h277 likely relate to the simulation's MW-like, yet cosmologically rare, early formation history and absence of late-time mergers.
This is coupled with h277's relatively low SFR, $\approx 1 \Msun$ yr$^{-1}$, over the last 6 Gyr, which is broadly consistent with observational estimates of the MW's star-formation history \citep{Smith12, Ruiz-Lara20}, because star formation and its associated feedback drive turbulence in the ISM, increasing $\sigma$ in gas.
The FIRE-2 galaxies we analyze have an average SFR today of $\approx 5-10 \Msun$ yr$^{-1}$ \citep{Gandhi22}.
\citep[See also Figure~9 of][for a comparison of SFRs in m12m, m12f, and h277]{Iyer20}.
Indeed, among the sample of cosmological zoom-in simulations we discuss above, most have higher SFRs than the MW and h277: for example, 60$\%$ of Auriga galaxies have SFRs of 5–10 $\Msun$ yr$^{-1}$, though individual galaxies occasionally have 1–2 $\Msun$ yr$^{-1}$.
Few simulations exhibit such low SFR over a long ($\gtrsim 6 \Gyr$) timescale, potentially indicating that this may be necessary to reproduce the MW's $\sigma(\tau)$.
Reinforcing this idea is that the one PHANGS galaxy with MW-like kinematics, NGC 1433, has a recent SFR of $1.1 \Msun {\rm yr}^{-1}$, the lowest among the PHANGS-MUSSE sample.

We thus conclude that, based only on observational comparisons, the MW is a kinematic outlier with unusually low $\sigma(\tau) / v_{\phi,0}$ among this sample of 19 observed nearby galaxies.
As such, it may not be surprising that among the sample of 59 cosmological zoom-in simulations with published $\sigma(\tau)$, only one (h277) matches the MW, while most simulations reasonably match the much larger sample of M31, M33, and (at least for stars older than $\approx 500 \Myr$) PHANGS.
This highlights the potential dangers in benchmarking simulations only against the MW, but this also highlights the remarkable value of ongoing measurements of stellar kinematics and their histories within nearby galaxies.

%Specifically, the early formation of h277's disk may have facilitated its consistent, low SFR at later times, because disk formation is a necessary prerequisite for disk settling, in which gaseous disks self regulate and correspondingly evolve towards colder kinematics \citep[i.e.,][]{Hopkins23}. h277's early disk formation also means that stars formed with rotationally-dominated kinematics for much of the galaxy's lifetime, and were subsequently heated within the [self-limiting geometric] confines of a disk. The absence of late-time mergers further allowed h277 to continue its cold, quiet evolution unabated/without disruption. Although this picture is indeed speculative, it provides a feasible explanation for the lower $\sigma(\tau)$ observed in h277 compared to other zoom-in simulations. It is perhaps more speculative -- though certainly tempting -- to view the MW's atypically cold kinematics as resulting from similar drivers. Indeed, while many aspects of the MW's formation remain uncertain, there is growing consensus that our Galaxy -- with its hectic youth and leisurely middle-age -- is not typical.[/representative.] [And indeed, an atypical galaxy may very well have atypically cold kinematics]. 

\section*{Acknowledgments}

F.M. and A.W. received support from: NASA, via FINESST grant 80NSSC24K1484; NSF, via CAREER award AST-2045928 and grant AST2107772; a Scialog Award from the Heising-Simons Foundation.
S.L.~acknowledges support from NSF grant AST-2109234 and HST grant AR-16624 from STScI.

\bibliography{main}{}
\bibliographystyle{aasjournal}

\appendix

\section{Comparing FIRE-2 Simulations and Observations without Scaling}
\label{a:unscaled}

\begin{figure*}
\centering
\includegraphics[width = \textwidth]{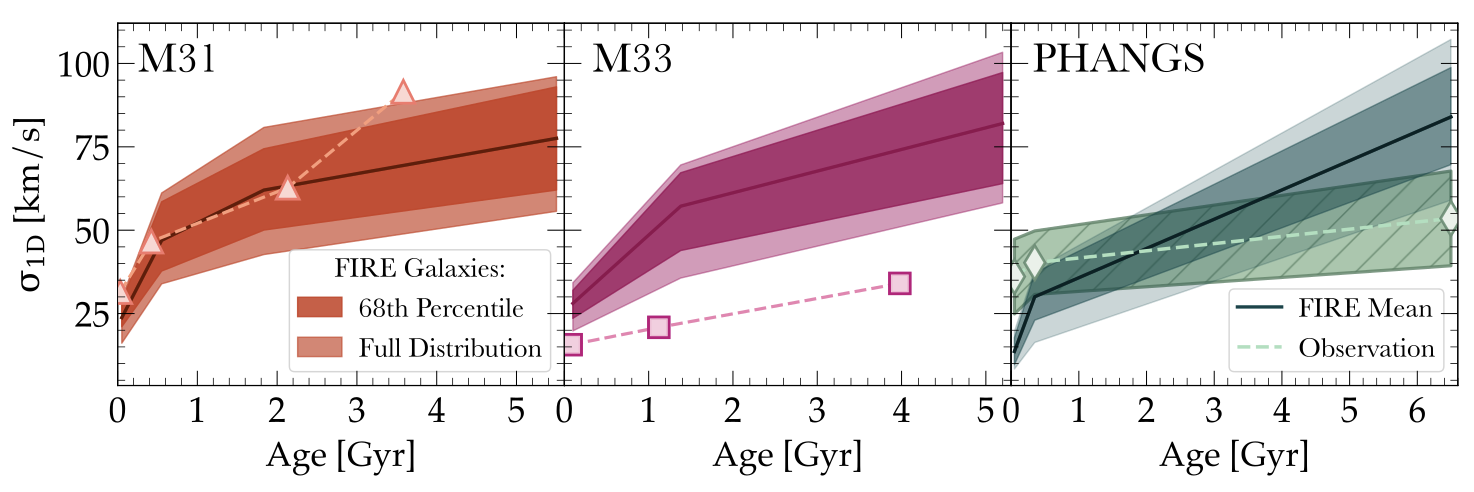}
\caption{
Same as Figure~\ref{fig:fire comp}, comparing FIRE-2 against M31, M33, and 16 galaxies from PHANGS, but \textit{without} scaling each galaxy's line-of-sight $\sigma$ to its rotational velocity for young stars today, $v_{\phi,0}$.
As a result, these $\sigma_{\rm 1D}$ are sensitive to the mass (gravitational potential) of the galaxy, in addition to its dynamical state.
In particular, $\sigma_{\rm 1D}$ for M33 is now much lower than FIRE-2.
However, given that M31 has a stellar mass more similar to most of these FIRE-2 galaxies, FIRE-2 still agrees with M31 about as well as in Figure~\ref{fig:fire comp}.
The typical galaxy in the PHANGS has slightly lower stellar mass than in our FIRE-2 sample, so its normalization is slightly lower relative to FIRE-2 than in Figure~\ref{fig:fire comp}.
We show these results for $\sigma_{\rm 1D}$ for completeness, but we consider the comparison against the dimensionless $\sigma_{\rm 1D} / v_{\phi,0}$ in Figure~\ref{fig:fire comp} more robust, because this normalizes out each galaxy's mass and therefore overall velocity scale.
}
\label{fig:unscaled}
\end{figure*}

In Section~\ref{subsec:fire comparison}, we compared $\sigma(\tau)$ in the FIRE-2 simulations against observed galaxies, scaling each galaxy by its rotational velocity, $v_{\phi,0}$, such that we compare $\sigma(\tau) / v_{\phi,0}$.
This dimensionless metric provides a fairer comparison of galaxy samples that span a factor of 10 in stellar mass.
However, many studies examine only $\sigma(\tau)$.
Therefore, here we show versions of Figure~\ref{fig:unscaled} and Figure~\ref{fig:mw fire} unscaled by $v_{\phi,0}$.
Note that the scaled versus unscaled versions differ only in normalization, not in shape, because $v_{\phi,0}$ is rotational velocity today and does not vary with age.
The results are qualitatively similar, because these FIRE-2 galaxies have stellar masses (and thus rotational velocities) similar to the MW, M31, and the median mass of the PHANGS-MUSE sample.
The primary difference is for M33, given its much lower stellar mass.

\begin{figure}
\centering
\includegraphics[width = 0.75\textwidth]{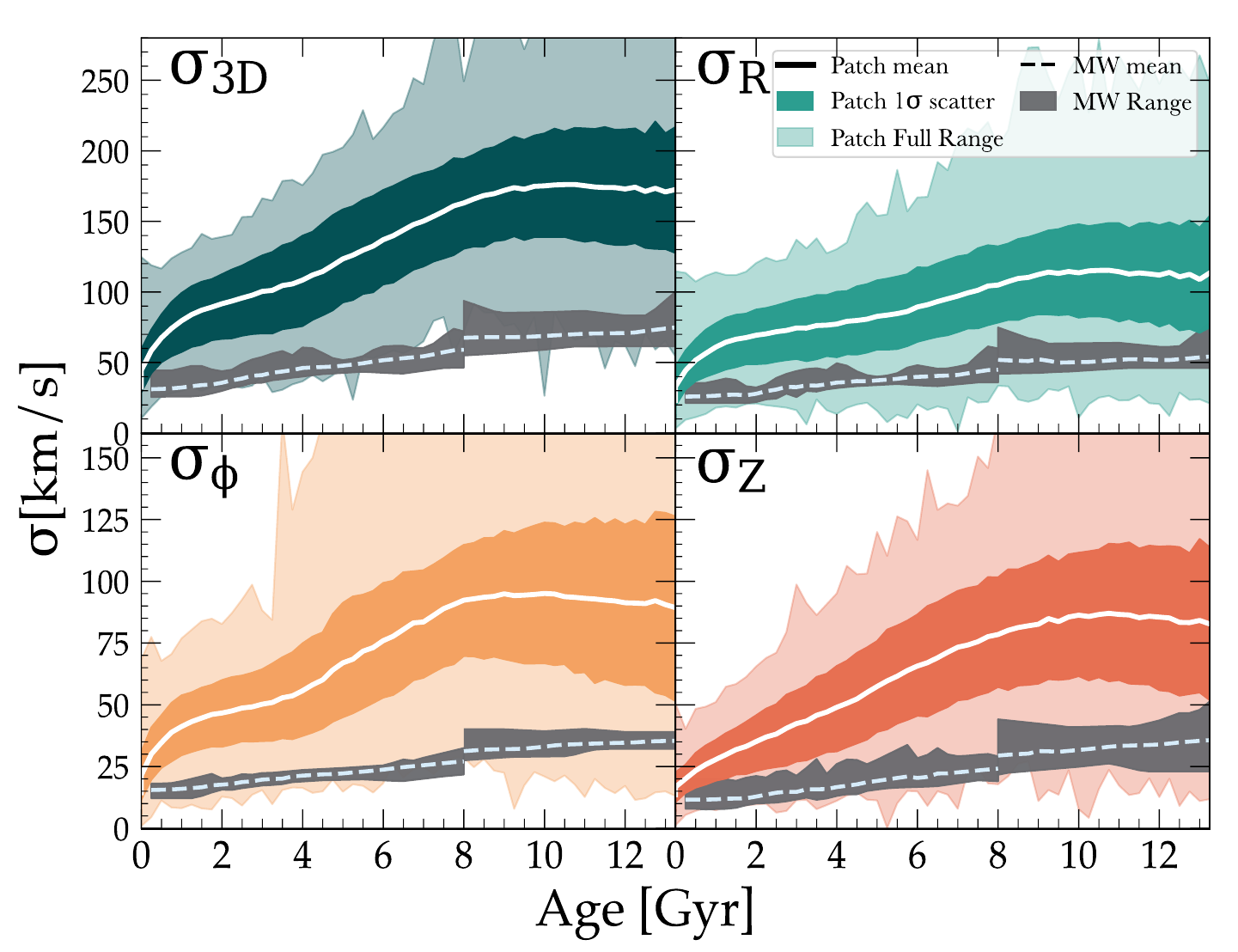}
\vspace{-2 mm}
\caption{
Same as Figure~\ref{fig:mw fire}, comparing FIRE-2 and the MW, but \textit{without} scaling $\sigma$ to each galaxy's rotational velocity today, $v_{\phi,0}$.
This yields nearly identical results: $\sigma$ in FIRE-2 simulations are typically larger than those measured in the MW, but they do agree for rare apertures.
This negligble difference when not scaling by $v_{\phi,0}$ arises because the median (or mean) $v_{\phi,0}$ of our FIRE-2 sample ($\approx 240 \kms$) is nearly identical to the value we use for MW ($238 \kms$).
}
\label{fig:mw fire unscaled}
\end{figure}

\section{Observations of Nearby Galaxies}
\label{a:external_obs}

\subsection{Observations of M31}
\label{a:external_obs_m31}

We now discuss the observations of \citet{Dorman2015} (\citetalias{Dorman2015}) in detail.
\citetalias{Dorman2015} measured the line-of-sight $\sigma(\tau)$ relation across the northern half of M31's disk using a sample of $\approx 5800$ individual stars with combined SPLASH spectroscopy and PHAT photometry. 
The photometric accuracy and precision afforded by deep HST observations allow clean color-magnitude-based separation of RGB, AGB, and MS stars. 
In turn, \citetalias{Dorman2015} split stars from their sample into 4 broad age bins based on their position on the optical CMD; these age bins have distinct mean ages (0.03, 0.4, 2, and 4 Gyr) despite their overlapping age distributions.
For each star, \citetalias{Dorman2015} first measured $\sigma$ in a surrounding circular patch of radius $\approx 760 \pc$. If the patch contained at least 15 stars in the same age bin, they calculated $\sigma$ from the second moment of the velocity distribution (of same-aged stars) using a kernel density estimator, while patches with fewer than 15 same-aged neighbors were ignored. The final reported $\sigma$ is the mean of all aperture-based $\sigma$ in each age bin.

\subsection{Observations of M33}
\label{a:external_obs_m33}

Similarly, we now discuss the observations of M33 by \citet{Quirk22} (\citetalias{Quirk22}) in more depth. The procedure of \citetalias{Quirk22} largely followed that of \citetalias{Dorman2015}, although their exact parameters necessarily differed. Specifically, \citetalias{Quirk22} divided their sample into 3 age bins with corresponding mean ages of $\approx 0.08$, 1, and 4 Gyr. As in \citetalias{Dorman2015}, the age distributions within each bin are broad and overlap, having 16th–84th percentile ranges of $\approx$ 20 – 180 Myr, 0.56-2.2 Gyr, and 1.4-9 Gyr. Furthermore, the aperture/patch radius over which \citetalias{Quirk22} measured $\sigma$ differed between age groups, with median values of $\approx 350$, 500, and 420 pc for the young, intermediate, and old populations, respectively.

Crucially, \citetalias{Quirk22} presented two $\sigma(\tau)$ relations for M33: one in which they removed ``halo" stars, defined as stars whose $v_{\rm LOS}$ are not consistent with belonging to a dynamically cold component, and one in which they did not. Although this dynamically hotter component only accounts for $\approx 14\%$ and 23\% of the initial intermediate-age and old-age populations, respectively, its inclusion significantly impacts the measured $\sigma$.
When the hotter component is excluded, $\sigma(\tau)$ is \textit{flat} with age, with all age bins having $\sigma \approx 16 \kms$.
However, including the hotter component resulted in $\sigma(\tau)$ that increases monotonically with age.
Specifically, $\sigma$ is unchanged for the youngest age bin, while the intermediate and oldest bins have $\sigma$ of $\approx 21$ and 35 km/s -- nearly 1.4 and 2 times larger than their ``uncontaminated'' values.
Nevertheless, given that neither \citetalias{Dorman2015} nor \cite{Pessa23} made similar dynamical cuts for M31 and PHANGS galaxies, we use the measurements from \citetalias{Quirk22} that do not remove dynamically-hot stars for consistency. This choice is further justified by the results of \cite{Beasley15}, who found a monotonically rising $\sigma(\tau)$ relation in M33 using observations of $\approx 80$ star clusters. Interestingly, this cluster-based relation exhibited a higher normalization and a stronger age dependence than the ``halo-contaminated'' relation of \citetalias{Quirk22}. We emphasize that regardless of the sample used, $\sigma(\tau) / v_{\phi,0}$ is always $\gtrsim2\times$ larger in M33 than in the MW.

\subsection{PHANGS-MUSE Survey}
\label{a:external_obs_phangs}

We now discuss -- in greater detail -- the PHANGS-MUSE sample and the procedure that \citet{Pessa23} (\citetalias{Pessa23}) used to measure $\sigma(\tau)$ in 19 galaxies. 
The PHANGS-MUSE sample lies within $\lesssim 20 \Mpc$ with low to moderate inclinations ($< 60^{\circ}$) and stellar masses of $\log(M_\star / M_\odot) = 9.4 - 11.0$, with a median of $\log(M_\star / M_\odot) = 10.5$.

As we discuss, one way to better compare the kinematics of different galaxies is to scale the dispersion by the galaxy's rotational velocity, \vphio. Although \citetalias{Pessa23} does not measure $v_{\phi, 0}$, earlier work by \cite{Lang20} provided rotation curves for 67 galaxies in the larger PHANGS-ALMA sample using high-resolution CO (2-1) maps. By parameterizing these rotation curves with smooth analytic fits, \cite{Lang20} determined the maximum (asymptotic) rotational velocities; however, three of the PHANGS-MUSE galaxies (NGC 2835, NGC 5068, and IC 5332) lacked robust smooth fits, leaving their maximum rotational velocities unreported. Consequently, our analysis focuses on the remaining 16 PHANGS-MUSE galaxies, which we hereafter refer to as the ``PHANGS galaxies/sample''.

Unlike the previously mentioned observations, these measurements are for stellar populations across $\approx 100 \pc$ scales, not individual stars. To extract stellar properties and kinematics, \citetalias{Pessa23} used the popular Penalized PiXel-Fitting (PPXf) code introduced by \cite{Cappellari04} and significantly upgraded by \citet{Cappellari17}. To measure stellar kinematics, they fit the observed Vornoi-binned spectrum to stellar templates (here the E-MILES templates of \cite{Vazdekis12} and their young extensions from \cite{Asa'd17}) convolved with a line-of-sight velocity distribution parameterized by a Gauss-Hermite expansion assuming a single stellar kinematic component.

The high spatial resolution and wide coverage of the PHANGS survey allowed \citetalias{Pessa23} to study stellar populations across multiple galactic environments separately. To more fairly compare different galaxies and account for radial and environmental trends, the $\sigma(\tau)$ relations in \citetalias{Pessa23} included only disk stars within $0.25 - 0.30 R_{\rm 25}$, where $R_{\rm 25}$ is the 25th magnitude isophotal B-band radius. For this sample, $R_{25}$ ranges from $7 - 34 \kpc$ with a mean value of $\approx 15 \kpc$, such that $0.25 - 0.30 R_{25}$ corresponds to $\approx 3.75 - 4.25 \kpc$. After making this geometric selection, \citetalias{Pessa23} calculated the mean for populations with luminosity-weighted ages within three age bins: young ($< 100 \Myr$), intermediate ($100 - 600 \Myr$), and old ($\gtrsim 3 \Gyr$).

\subsection{Observations of the Milky Way}
\label{a:mw}

Finally, we describe the 13 works that we compiled that measured $\sigma(\tau)$ of the MW.
\cite{Nordstrom2004} introduced the Geneva-Copenhagen Survey (GCS). The GCS combined Stromgren photometry and radial velocity measurements with Hipparcos proper motions and parallaxes to measure late-type stars in the Solar neighborhood, ultimately creating an all-sky, kinematically-unbiased catalog of $\approx 16,500$ F and G stars that is volume complete within 40 pc of the Sun.

Since \cite{Nordstrom2004}, the GCS catalog has undergone multiple revisions. First, \cite{Holmberg07} used new temperature and metallicity calibrations to update the metallicites and ages. Second, \cite{Holmberg09} implemented newly revised Hipparcos parallaxes to significantly improve distance measurements, resulting in improved kinematics and ages as well. Lastly, \cite{Casagrande11} adopted a corrected effective temperature scale to provide more accurate and precise metallicities and ages. These updated ages are generally $\approx 1 \Gyr$ younger than their earlier values, with the oldest ages often decreasing by $\approx 3 - 6 \Gyr$.
However, a substantial number of stars previously deemed young (age $< 1 \Gyr$) now have ages ranging between $0.5 - 10 \Gyr$ \citep[see Figure~A2 of][]{Casagrande11}. Despite these changes, all of these works show similar $\sigma(\tau)$. We include \cite{Nordstrom2004}, \cite{Holmberg09}, and \cite{Casagrande11} in our compilation, given their prevalence in the literature, and to show how different age determinations impact the measured $\sigma(\tau)$ for the same stellar sample.

All of the GCS analyses used the Galactic coordinate system and a heliocentric Cartesian system, so they quoted $\sigma$ in Galactic space-velocities (U,V,W), where U points towards the Galactic center, V the direction of rotation, and W the north Galactic pole.
However, with the exception of \cite{Silva18}, the other works (including our own) used the Galactocentric coordinate system ($R$, $\phi$, $Z$).
We tested how using the Galactic coordinate system impacts the $\sigma$ measured in FIRE-2, and we found that for aperture-based measurements, the median vertical (W) and 3D $\sigma$ were generally unchanged, as expected.
However, the median $\sigma$ in U is typically $\approx 15 - 20\%$ \textit{lower} than $\sigma$ in $v_R$, while V is typically $\approx 20 - 25\%$ \textit{higher} than $\sigma$ in $v_{\phi}$, across all ages.
Although the in-plane components of $\sigma$ are not exactly analogous for the two coordinate systems, we show the U and V velocity dispersions as $\sigma_{R}$ and $\sigma_{\phi}$ in Figure~\ref{fig:mw obs} for consistency.

We know discuss works that measured $\sigma(\tau)$ in the Gaia era, grouping similar works together rather than adopting a purely chronological ordering. 
\cite{Yu_18} analyzed a sample of $\approx 3,500$ SGB/RGB stars that have LAMOST spectroscopy and Gaia DR1 astrometry (stars in the Tycho-Gaia Astrometric Solution catalog). Following the likelihood technique of \citet{Liu15}, \cite{Yu_18} estimated stellar ages by comparing the observed stellar parameters with theoretical isochrones, leading to typical age uncertainties of $\approx 30\%$. This sample is fairly local: requiring reliable astrometric measurements limits the sample to within 1 kpc of the solar neighborhood. Furthermore, \cite{Yu_18} divided their sample into two equal populations based on the stars' distance from the mid-plane: $\approx 1780$ stars with $|Z| < 270 \pc$ and 1780 with $|Z| > 270 \pc$.
Both populations exhibit similar 3D and radial $\sigma$, with the population at $|Z| < 270 \pc$ having a larger azimuthal $\sigma$ and a smaller vertical $\sigma$ at most ages. For brevity, we include only the population at $|Z| < 270 \pc$.

\cite{Kordopatis23} derived stellar parameters, including orbital properties, masses, and isochrone ages, for $\approx 5$ million stars with radial velocities from Gaia-EDR3, by combining Gaia-EDR3 spectroscopy, photometry, and astrometry with 2MASS infrared photometry. \cite{Kordopatis23} presented $\sigma(\tau)$ for RGB and MSTO stars within 1 kpc of the Sun.
They claim that their stellar ages are most reliable for MSTO stars, with all but the least-massive MSTO stars having age errors $\lesssim 50\%$. Therefore, we include only their MSTO data, sampling their trendline every 0.5 Gyr in Figure~\ref{fig:mw obs}.

\cite{Sharma21} compared the results of multiple complementary surveys, including GALAH, LAMOST, APOGEE, Kepler, and Gaia, to explore how the $\sigma(\tau)$ depends on age, angular momentum, metallicity, and $|Z|$.
However, \cite{Sharma21} mainly presented (dimensionless) normalized $\sigma$, in part to compare the relative profiles of these different works.
That said, they presented raw $\sigma$ from GALAH DR3 for $\approx 100,000$ MSTO stars in the extended solar neighborhood, defined via $|R-R_\odot|$ and $|Z|$, which we sample every 0.5 Gyr.
The GALAH+ DR3 catalog \citep{Buder21} provides stellar ages estimated using the Bayesian Stellar Parameter Estimation code from \cite{Sharma18}, which provides a Bayesian estimate of stellar parameters from (PARSEC-COLIBRI) isochrones. When we adopt the same quality cuts as \cite{Sharma21}, we find an average age uncertainty of 16.7\%. 
Interestingly, \cite{Sharma21} compared $\sigma(\tau)$ of RGB and MSTO stars in GALAH DR3 and LAMOST DR4, finding that while RGB stars from both surveys and MSTO stars in LAMOST had $\sigma$ that significantly flattened or even decreased with age above 8 Gyrs, $\sigma$ of MSTO stars from GALAH continued to increase monotonically with age.
The authors posited that this behavior may stem from the lower age uncertainties of the GALAH MSTO sample.

Also using GALAH DR3 results, \cite{Hayden22} derived new ages for $\approx 215,000$ stars using their measured elemental abundances and overall metallicities. They quote typical age uncertainties for their abundance-based ages of $\approx 1 - 2 \Gyr$.
However, unlike for other methods, these uncertainties are largely age-independent, so older stars have more accurate ages than younger stars. \cite{Hayden22} showed $\sigma(\tau)$ for stars at $R = 7.1 - 9.1 \kpc$ and $|Z| < 0.5 \kpc$.
We sample the $\sigma$ of their ``All GALAH Stars'' population every 0.5 Gyr.

\cite{Silva18} combined optical and infrared photometry, APOGEE spectroscopy, and Kepler astroseismology to construct a photometrically complete sample of $\approx 1200$ stars across 2 kpc of the solar annulus. The sample has a median age uncertainty of 28.5\%. \cite{Silva18} also used the Galactic coordinate system.

\cite{Miglio21} and \cite{Lagarde21} considered overlapping samples of red-giant-branch (RGB) stars with reliable constraints from Kepler astroseismology, APOGEE spectroscopy, and Gaia astrometry.
However, each work further refined their sample and determined stellar properties, in particular stellar ages, differently. \cite{Miglio21} selected only giant stars with robust mass measurements, producing a final sample of $\approx 3,300$ stars with a median random age uncertainty of 23\%. From this sample, \cite{Lagarde21} further selected for stars in the APOKASC catalog, yielding a final sample of $\approx 2,800$ giant stars. Both works divided their sample into high- and low-$\alpha$ populations, with \cite{Lagarde21} further dividing their low-$\alpha$ population by [Fe/H]. We show both populations from \cite{Miglio21}, and we show the low-$\alpha$ and the metal-poor high-$\alpha$ populations from \cite{Lagarde21}, omitting their metal-rich high-$\alpha$ population for brevity, given that its $\sigma$ lie between those of the other populations.

As in \cite{Yu_18}, \cite{Sun24} also combined LAMOST and Gaia observations.
However, the improved proper motions that Gaia DR2 and EDR3 provided allow them to measure red clump stars throughout much of the disk, with a sample of $\approx 130,000$ stars spanning $R = 6 - 15 \kpc$ and $|Z| < 3 - 4 \kpc$.
They determined stellar ages from a machine-learning model trained on a sample of thousands of red clump stars in the LAMOST-KEPLER field with accurate astroseismic ages (following \citet{Huang18}), leading to a typical age uncertainty of $\approx 30\%$.
In their analysis, \cite{Sun24} subdivided their entire sample into various spatial and population bins, providing 3D kinematic maps and $\sigma(\tau)$ for each.
For this comparison, we use their ``thin'' and ``thick'' disk measurements for stars within $R = 7 - 9 \kpc$ and $|Z| < 1 \kpc$ (shown in the bottom-left panels of their Figure 9), where they determined the low-$\alpha$ and high-$\alpha$ disk populations using an empirical cut in the [Fe/H]–[$\alpha$/Fe] plane.

Similarly, \cite{Mackereth19} also provided a comprehensive dissection of the low- and high-$\alpha$ disks using stars with spectra from APOGEE DR14 and ages determined using machine learning models trained on APOGEE stars that have robust asteroseismic ages from Kepler. As in \cite{Sun24}, the improved astrometry of Gaia DR2 allowed \cite{Mackereth19} to probe stars throughout a large volume of the disk: $R = 4 - 13 \kpc$ and $|Z| < 2 \kpc$. \cite{Mackereth19} presented $\sigma(\tau)$ for stars across $R = 6 - 11 \kpc$.
We make no further radial selection and include their data for all mean radii, because we cannot determine these stars' \textit{current} radii from their \textit{mean} radii. However, low- and high-$\alpha$ populations generally have slightly different radial ranges, with the low-$\alpha$ data spanned the entire radial range, while the high-$\alpha$ data had mean radii limited to $R \approx 6 - 7 \kpc$. Interestingly, $\sigma_R$ showed little dependence on the mean orbital radius, while $\sigma_Z$ increased with increasing mean radius at a fixed age. The final sample of $\approx 66,000$ stars has a median age uncertainty $\approx 30-35\%$, though the authors note that for stars older than 10 Gyr their ages are likely to be underestimated and subject to large uncertainties.

Recently, \cite{Anders23} provided a catalog with precise stellar ages for $\approx 180,000$ RGB stars from APOGEE.
They used a supervised machine learning method trained on $\approx 3,000$ RGB stars with existing ages from both APOGEE spectra and Kepler asteroseismic ages. They claim that the catalog has a median age uncertainty of 17\%. Adopting distances and velocities from the StarHorse value-added catalog from APOGEE DR17 \citep{Queiroz23}, which combined APOGEE DR17 spectra, broad-band photometric data, and Gaia EDR3 parallaxes to derive astrophysical parameters for APOGEE stars, \cite{Anders23} presented $\sigma(\tau)$ for stars with $|Z| < 1 \kpc$ across multiple $R$. To maintain consistency with the other MW observations, we present their data for $R = 7 - 9 \kpc$.



\end{document}